\PassOptionsToPackage{unicode}{hyperref}
\PassOptionsToPackage{hyphens}{url}
\PassOptionsToPackage{dvipsnames,svgnames,x11names}{xcolor}
\documentclass[
  12pt]{article}

\usepackage{amsmath,amssymb}
\usepackage{iftex}
\ifPDFTeX
  \usepackage[T1]{fontenc}
  \usepackage[utf8]{inputenc}
  \usepackage{textcomp} % provide euro and other symbols
\else % if luatex or xetex
  \usepackage{unicode-math}
  \defaultfontfeatures{Scale=MatchLowercase}
  \defaultfontfeatures[\rmfamily]{Ligatures=TeX,Scale=1}
\fi
\usepackage{lmodern}
\ifPDFTeX\else  
\fi
\IfFileExists{upquote.sty}{\usepackage{upquote}}{}
\IfFileExists{microtype.sty}{% use microtype if available
  \usepackage[]{microtype}
  \UseMicrotypeSet[protrusion]{basicmath} % disable protrusion for tt fonts
}{}
\makeatletter
\@ifundefined{KOMAClassName}{% if non-KOMA class
  \IfFileExists{parskip.sty}{%
    \usepackage{parskip}
  }{% else
    \setlength{\parindent}{0pt}
    \setlength{\parskip}{6pt plus 2pt minus 1pt}}
}{% if KOMA class
  \KOMAoptions{parskip=half}}
\makeatother
\usepackage{xcolor}
\makeatletter
\ifx\paragraph\undefined\else
  \let\oldparagraph\paragraph
  \renewcommand{\paragraph}{
    \@ifstar
      \xxxParagraphStar
      \xxxParagraphNoStar
  }
  \newcommand{\xxxParagraphStar}[1]{\oldparagraph*{#1}\mbox{}}
  \newcommand{\xxxParagraphNoStar}[1]{\oldparagraph{#1}\mbox{}}
\fi
\ifx\subparagraph\undefined\else
  \let\oldsubparagraph\subparagraph
  \renewcommand{\subparagraph}{
    \@ifstar
      \xxxSubParagraphStar
      \xxxSubParagraphNoStar
  }
  \newcommand{\xxxSubParagraphStar}[1]{\oldsubparagraph*{#1}\mbox{}}
  \newcommand{\xxxSubParagraphNoStar}[1]{\oldsubparagraph{#1}\mbox{}}
\fi
\makeatother

\providecommand{\tightlist}{%
  \setlength{\itemsep}{0pt}\setlength{\parskip}{0pt}}\usepackage{longtable,booktabs,array}
\usepackage{calc} % for calculating minipage widths
\usepackage{etoolbox}
\makeatletter
\patchcmd\longtable{\par}{\if@noskipsec\mbox{}\fi\par}{}{}
\makeatother
\IfFileExists{footnotehyper.sty}{\usepackage{footnotehyper}}{\usepackage{footnote}}
\makesavenoteenv{longtable}
\usepackage{graphicx}
\makeatletter
\def\maxwidth{\ifdim\Gin@nat@width>\linewidth\linewidth\else\Gin@nat@width\fi}
\def\maxheight{\ifdim\Gin@nat@height>\textheight\textheight\else\Gin@nat@height\fi}
\makeatother
\setkeys{Gin}{width=\maxwidth,height=\maxheight,keepaspectratio}
\makeatletter
\def\fps@figure{htbp}
\makeatother

\makeatletter
\@ifpackageloaded{caption}{}{\usepackage{caption}}
\AtBeginDocument{%
\ifdefined\contentsname
  \renewcommand*\contentsname{Table of contents}
\else
  \newcommand\contentsname{Table of contents}
\fi
\ifdefined\listfigurename
  \renewcommand*\listfigurename{List of Figures}
\else
  \newcommand\listfigurename{List of Figures}
\fi
\ifdefined\listtablename
  \renewcommand*\listtablename{List of Tables}
\else
  \newcommand\listtablename{List of Tables}
\fi
\ifdefined\figurename
  \renewcommand*\figurename{Figure}
\else
  \newcommand\figurename{Figure}
\fi
\ifdefined\tablename
  \renewcommand*\tablename{Table}
\else
  \newcommand\tablename{Table}
\fi
}
\@ifpackageloaded{float}{}{\usepackage{float}}
\floatstyle{ruled}
\@ifundefined{c@chapter}{\newfloat{codelisting}{h}{lop}}{\newfloat{codelisting}{h}{lop}[chapter]}
\floatname{codelisting}{Listing}

\makeatother
\makeatletter
\@ifpackageloaded{caption}{}{\usepackage{caption}}
\@ifpackageloaded{subcaption}{}{\usepackage{subcaption}}
\makeatother

\ifLuaTeX
  \usepackage{selnolig}  % disable illegal ligatures
\fi
\usepackage[]{natbib}
\usepackage{bookmark}

\IfFileExists{xurl.sty}{\usepackage{xurl}}{} % add URL line breaks if available
\hypersetup{
  pdftitle={The Noisy Work of Uncertainty Visualisation},
  pdfauthor={Harriet Mason; Dianne Cook; Sarah Goodwin; Emi Tanaka; Susan VanderPlas},
  pdfkeywords={data visualisation, statistical graphics, data
science, information visualisation},
  colorlinks=true,
  linkcolor={blue},
  filecolor={Maroon},
  citecolor={Blue},
  urlcolor={Blue},
  pdfcreator={LaTeX via pandoc}}

\begin{document}

\def\spacingset#1{\renewcommand{\baselinestretch}%
{#1}\small\normalsize} \spacingset{1}

%%%%%%%%%%%%%%%%%%%%%%%%%%%%%%%%%%%%%%%%%%%%%%%%%%%%%%%%%%%%%%%%%%%%%%%%%%%%%%

\date{June 11, 2026}
\title{\bf The Noisy Work of Uncertainty Visualisation}
\author{
Harriet Mason\\
Department of Econometrics and Business Statistics, Monash University\\
and\\Dianne Cook\\
Department of Econometrics and Business Statistics, Monash University\\
and\\Sarah Goodwin\\
Department of Human Centred Computing, Monash University\\
and\\Emi Tanaka\\
Biological Data Science Institute, The Australian National University\\
and\\Susan VanderPlas\\
}
\maketitle

\bigskip
\bigskip
\begin{abstract}
Better representation of the uncertainty in a data visualisation is a
focus of recent research activity. A problem with the current literature
is that there is a lack of clarity about the definition of uncertainty
and what it means to represent it in a plot. This confusion results in a
significant amount of conflicting results in the literature, especially
in experiments that assess the effectiveness of different uncertainty
representations. In this review, we summarise the current literature,
provide workable definitions, and illustrate these definitions with
examples. In doing so, we ask what it really takes to achieve
transparency in statistical graphics. It is hoped that it will be useful
for guiding new graphics methodology and experimental research.
\end{abstract}

\noindent%
\vfill

\newpage
\spacingset{1.9} % DON'T change the spacing!

\section{Introduction}\label{introduction}

What do we mean when we talk about ``uncertainty visualisation''? The
phrase can feel contradictory to anyone familiar with the term. Among
statisticians, ``uncertainty'' is often discussed as an omnipresent
spectre touching every stage of our analysis without ever being fully
seen. Authors will often mention that the phrase is vague
\citep{Spiegelhalter2017, Griethe2006}, or avoid defining it by
describing a list of things uncertainty \emph{could} be
\citep{Kinkeldey2014, Hullman2016}, but rarely do authors attempt to
discuss what uncertainty actually \emph{is}. By contrast, visual
statistics (information visualisations, data plots) are one of the most
powerful tools in the statistician's toolbox, allowing for quick and
memorable communication that identifies quirks in our data that we
didn't even know to look for. We see this in datasets such as Anscombe's
quartet \citep{anscombe} or the Datasaurus Dozen
\citep{datasaurus, datasaurpkg}, where visual statistics are able to
highlight elements of the data that are invisible to the typical summary
statistics. We also see this in recall experiments, where simply
sketching a distribution before recalling statistics or making
predictions can greatly increase the accuracy of those measures
\citep{Hullman2018, Goldstein2014}. Taken together, uncertainty
visualisation implies a need to pull back the curtain and explore the
unknowns of our analysis.

As nice as this sentiment is, it turns out to be easier said than done.
Reviews on uncertainty visualisation rarely offer tried and tested rules
for effective uncertainty visualisation, instead commenting on the
difficulties faced when trying to summarise the field.
\citet{Kinkeldey2014} found most experimental methods to be ad hoc, with
no commonly agreed upon methodology, formalisations, or a greater goal
of describing general principles. \citet{Hullman2016} noticed there is a
serious noise issue in the field, with noise from participants
misunderstanding visualisations, misinterpreting questions, and
incorrectly applying heuristics, overwhelming any information we can
glean from studies. \citet{MacEachren2005} identified so much
contradicting evidence that they spent an entire page discussing the
conflicting evidence for the question ``Should I map uncertainty to
colour hue?'' \citet{Spiegelhalter2017} concluded that different plots
are good for different things, arguing against a universal best plot for
all people and circumstances. \citet{uncertchap2022} summarised several
cognitive effects that repeatedly arise in uncertainty visualisation
experiments; these effects were each discussed in isolation as a list of
considerations rather than an overarching theory for effective
uncertainty visualisation.

\begin{quote}
``Science is built up of facts, as a house is built of stones; but an
accumulation of facts is no more a science than a heap of stones is a
house.'' - Henri Poincaré (1905)
\end{quote}

While these reviews are thorough in scope, none discuss how the existing
literature contributes to the broader goal of uncertainty visualisation
-- that is, despite the wealth of reviews, the field of uncertainty
visualisation remains a heap of stones. There is a mountain of work that
identifies common heuristics found in uncertainty visualisations,
evaluates competing plot designs, or starts a theoretical discussion on
a niche aspect of the field. While important, each of these papers
offers up its own bespoke motivation and methodology, with little
reference to the uncertainty visualisation papers outside their fiefdom.
The field is in desperate need of a unifying theory that can tie the
conflicting and siloed research together. This review attempts to
address this issue by offering a novel perspective on the uncertainty
visualisation problem. That is, we will use the wealth of established
stone to construct a foundation to build a house.

\section{The purpose of uncertainty
visualisation}\label{the-purpose-of-uncertainty-visualisation}

Mentions of ``uncertainty visualisation'' start springing up around
1990, across several different fields
\citep{Ibrekk1987, MacEachren1992}, each with its own motivation for the
work. In computer science, the area appears to be motivated by issues in
the public's perception of random variables, with the hope that
visualisations would give laypeople the ability to extract important
information from graphical representations \citep{Ibrekk1987}. With
similar concerns about the public's understanding of randomness, the
fields of psychology, statistics, and economics used ``uncertainty
visualisations'' as a communication tool to mitigate the psychological
bias associated with the communication of risk, a topic of concern since
the early 1980s \citep{Spiegelhalter2017}. In cartography, it was
motivated by the inherent uncertainty of geoscience data, the practical
use of visualisation as an exploratory tool, and the constrained visual
channels from map representations \citep{MacEachren1992}. These
disparate motivations have blended together, and today, uncertainty
visualisation is usually motivated by the vague goal of
``decision-making''. This term has been used to mean the mitigation of
psychological bias to ensure economically rational decisions
\citep{Padilla2022, Padilla2021, Kale2021}, the facilitation of trust or
confidence \citep{Zhao2023, Yang2023}, the ability to extract values
related to a distribution \citep{Sarma2023}, the prevention of false
discovery in plots \citep{Sarma2024, Koonchanok2023}, or the extraction
of some other metric that is vaguely related to ``uncertainty''
\citep{Chakraborty2024, Ndlovu2023}. This gradual scope creep of the
field, motivated by the hazy definition of ``decision-making'', is the
most likely culprit for the jumbled literature that makes up the field
today.

Given that there is so much subliminal disagreement in uncertainty
visualisation, how did these disparate motivations come to be seen as
interchangeable? All discussions on uncertainty visualisation seem to
have a common thread that connects them: the belief that the ultimate
goal of uncertainty visualisation is not trust, rationality, or value
extraction, but \emph{transparency}. We see it said directly in reviews
of the field \citep{uncertchap2022}, or when authors claim that failing
to include uncertainty is akin to fraud or lying
\citep{Hullman2020a, Manski2020}. We see it when authors assert that
uncertainty communicates the legitimacy (or illegitimacy) of the
conclusion drawn from visual inference \citep{Kale2018, Griethe2006}. We
see it when authors say uncertainty visualisations should communicate a
degree of confidence \citep{Correll2018, Boukhelifa2012} or validity
\citep{Hullman2020a, Griethe2006} in our conclusions. We see it when
authors suggest uncertainty visualisation should ``guide, qualify, or
soften our judgements of uncertain data'' (e.g. \citet{Leland2005} in
his seminal work on the grammar of graphics). These authors are not
wrong about the need for transparency in science communication: a
six-month survey of anti-mask groups on Facebook during the COVID-19
pandemic showed that anti-maskers made persuasive arguments by
exploiting inherent uncertainty ignored by pro-maskers \citep{Lee2021}.

Uncertainty visualisation is motivated by the need for a sort of
``visual hypothesis test'', a sentiment expressed by some authors
directly \citep{Correll2014, MacEachren1992}. A successful uncertainty
visualisation would act as a ``statistical hedge'' for any inference we
make using the graphic. Since the purpose of a visualisation is to give
a quick gist of the information \citep{Spiegelhalter2017}, this hedging
should be communicated visually without the need for complicated mental
calculations. Therefore, an effective uncertainty visualisation should
not just ``show'' uncertainty; untrustworthy conclusions should
\emph{not be visible}. If we refer to the conclusion we draw from a
graphic as its signal, where uncertainty should make this signal harder
to read as the ``noise'' increases, we can summarise the above
information into three key requirements. A good uncertainty
visualisation should:

\begin{enumerate}
\def\labelenumi{\arabic{enumi})}
\tightlist
\item
  Reinforce justified signals to encourage confidence in results.
\item
  Hide spurious signals that are overwhelmed by noise.
\item
  Perform tasks 1) and 2) in a way that is proportional to the level of
  confidence in those conclusions.
\end{enumerate}

Usually, visualisations that are unconcerned with uncertainty have no
issue showing justified signals, but struggle with the display of
unjustified signals. Therefore, we \textbf{suggest calling this approach
to uncertainty visualisation ``signal-suppression''} since it primarily
differentiates itself from the normal ``noiseless'' visualisation
approach through criterion (2). This is the main criterion we will use
to assess the current literature on uncertainty visualisation.

\section{Current Approaches}\label{current-approaches}

\subsection{Ignoring uncertainty}\label{ignoring-uncertainty}

The most common way to visualise uncertainty is to simply not. A study
conducted by \citet{Hullman2020a} found that only a quarter of authors
surveyed included uncertainty in 50\% or more of their visualisations,
in part because authors are not sure how to calculate uncertainty. This
is not entirely unreasonable, given that even visualisation authors
themselves seem to be in conflict about what exactly uncertainty is. We
will start with visualisations that ignore uncertainty with the hope
that by looking at where uncertainty isn't, we can better understand
where it is.

\subsubsection{What is uncertainty?}\label{what-is-uncertainty}

It is surprisingly hard to describe what uncertainty is. Most authors
avoid the problem and describe the many characteristics of uncertainty.
Often, uncertainty is split by factors such as whether it is due to true
randomness or a lack of knowledge
\citep{Begg2014, Spiegelhalter2017, Gustafson2019, uncertchap2022, Hullman2016, utypo};
quantifiable or unquantifiable
\citep{Spiegelhalter2017, utypo, uncertchap2022}; scientific or human
\citep{Benjamin2018, Gustafson2019}; systematic or random
\citep{Sanyal2009}; statistical or bounded
\citep{Gschwandtnei2016, Olston2002}; accuracy or precision
\citep{Griethe2006, Benjamin2018, Hullman2016}; etc. There are enough
qualitative descriptors of uncertainty to fill a paper, but none of this
is particularly helpful in understanding how to integrate it into a
visualisation.

Rather than trying to define uncertainty by looking at the myriad ways
in which it \emph{does} appear in an analysis, we may find it easier to
look at where it \emph{does not}. Descriptive statistics describe our
sample as it is and summarise large data into a usable format, but they
are not seen as the primary goal of modern statistics. In 19th-century
England, \emph{positivism} was the popular philosophical approach to
science (positivists included famous statisticians such as Francis
Galton and Karl Pearson). Practitioners of the approach believed
statistics ended with descriptive statistics, as science must be based
on actual experience and observations \citep{Otsuka2023}. In order to
make statements about population statistics, future values, or new
observations, we need to perform inference, which requires the
assumption of the ``uniformity of nature'', that is, we need to assume
that unobserved phenomena should be similar to observed phenomena
\citep{Otsuka2023}. Positivists believed referencing the unobservable
was bad science, embracing descriptive statistics due to the inherent
certainty associated with them. Since uncertainty is nonexistent in
descriptive statistics, it is clear that uncertainty is a by-product of
inference: uncertainty is the noise that is both inseparable from our
inference and meaningless without it.

Rather than extracting just one element of the distribution, if you can
retain the whole distribution, that not only allows the uncertainty
calculation to be reproduced, but also makes it possible to derive other
estimates as well. I think it's also important to acknowledge that if
you use this approach, you need to know the ``chicken''
(i.e.~distribution) that gave birth to the ``egg'' (uncertainty
estimate). In practice, uncertainties are sometimes calculated by other
people or organisations, and the process used to derive them may not be
known.

If we consider uncertainty to be a by-product of statistical inference,
then uncertainty visualisations are the plots that depict an estimate,
and therefore have an associated uncertainty. The most complete
description of these estimates is their distributions. Rather than
extracting just one element of the distribution, if you can retain the
whole distribution, that not only allows the uncertainty calculation to
be reproduced, but also makes it possible to derive other estimates as
well. Suggesting distributions as a representation of uncertainty is not
new. \citet{Kay2023} originally suggested thinking about uncertainty
visualisations as visualisations with distribution inputs, to replace
the commonly used mean and standard deviation, removing the assumption
of a Gaussian distribution. In practice, uncertainties are sometimes
calculated by other people or organisations, and the process used to
derive them may not be known. While some researchers believe these
abstract notions of uncertainty, such as credibility
\citep{Thomson2005}, forecaster confidence \citep{Padilla2021}, or
uncertainty about uncertainty \citep{Hadjimichael2024}, are too complex
to be quantified, this is not necessarily true. Abstract concepts such
as human belief or credibility are regularly quantified by Bayesians,
and hierarchical approaches are often used to model uncertainty about
uncertainty.

\subsubsection{Example: ignoring
uncertainty}\label{example-ignoring-uncertainty}

\begin{figure}[t]

\centering{

\includegraphics[width=1\textwidth,height=\textheight]{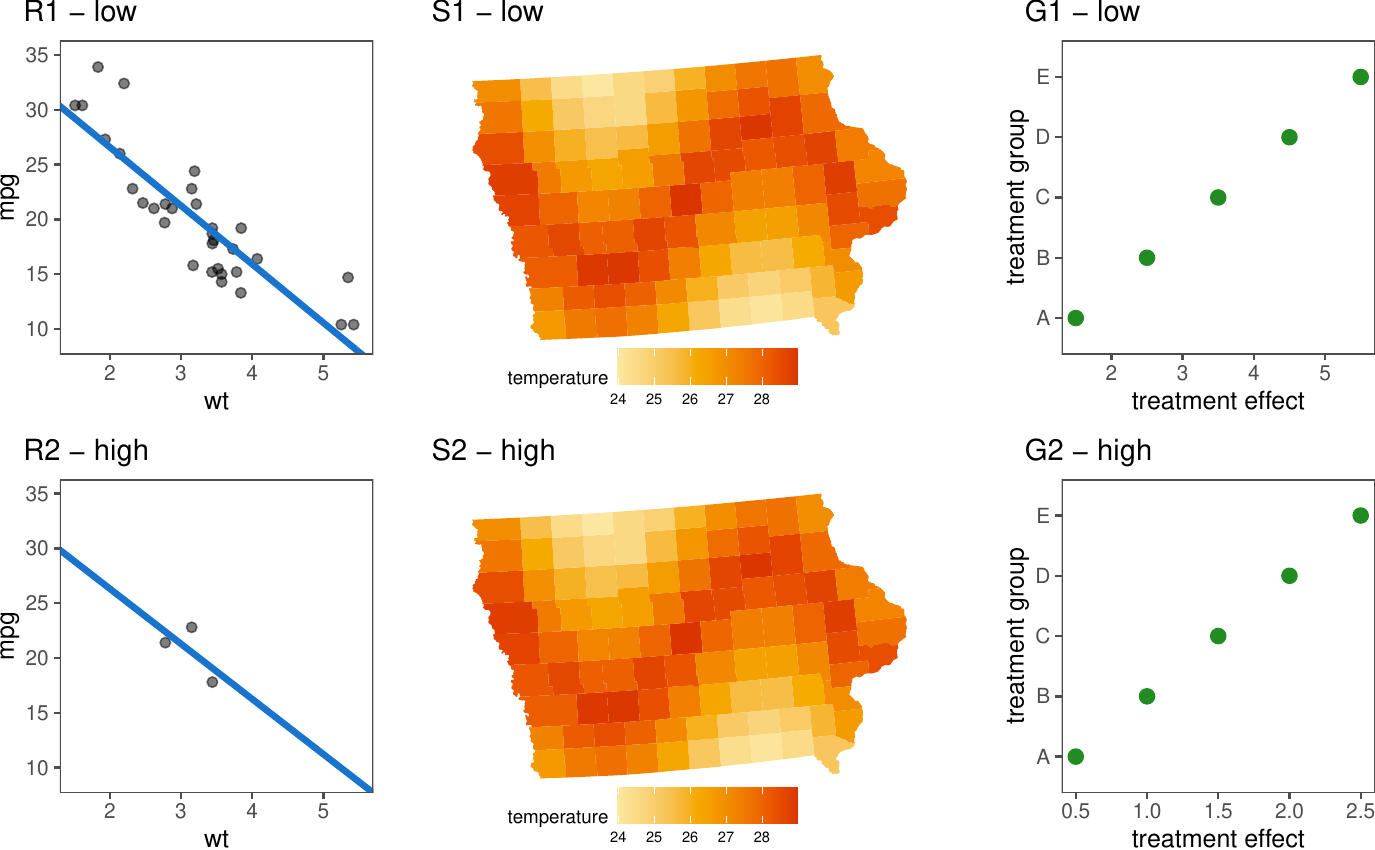}

}

\caption{\label{fig-ignore}Three example types of data and associated
plots that will be used to illustrate various choices of uncertainty
representation throughout the paper: scatterplot and regression line
(R1, R2), spatial choropleth map (S1, S2), grouped dotplot (G1, G2). Two
levels of uncertainty (low, high) are used with each example. Ignoring
the points in R1 and R2, there are no differences between the high and
low uncertainty versions. Ignoring uncertainty can lead to
misrepresentation of data.}

\end{figure}

If visualising uncertainty is fundamentally visualising a set of random
variables, what does ``ignoring'' uncertainty look like?
Figure~\ref{fig-ignore} shows plots of data from three different
scenarios, each with either high or low uncertainty. The plots show the
expected value of the input distribution. In plots R1 and R2, the
expected value is the line from a simple linear regression on a car's
miles per gallon (\texttt{mpg}) and weight (\texttt{wt}) using the
\texttt{mtcars} data, or a subset of it (available in \texttt{ggplot2}
\citep{ggplot2}, and originally from \citet{mtcars}), with differing
sample sizes creating differing levels of uncertainty. Plots S1 and S2
show a choropleth map of Iowa, where counties are coloured according to
a simulated temperature measurement, with measurement error being the
uncertainty associated with the measuring instrument. Plots G1 and G2
show samples simulated from five populations (A-E). In G1, distributions
have the same low variance, and in G2, they have the same high variance.
In the linear regression, we see a downward trend for both levels of
uncertainty. Here, we elected to plot the data under the fit, to provide
context for this example, and help follow the thinking through to the
later illustrations. Often, only the fitted line is shown. In the map,
you should see a sine wave spatial trend, and in the univariate
distributions, an incremental increase in treatment. If we were to ask a
reader, ``Can you see a difference between the plots in the top row
versus the bottom row? Is the strength of the trend communicated through
the visualisation?'' The answer to both of these questions will be no
for S1, S2, G1, and G2, as the high and low variance cases are
identical. For R1 and R2, the answer would likely only relate to the
points in the plots, not the regression line.

These examples will serve as touchstones for a discussion of uncertainty
visualisation, focusing on the approaches suggested in the literature.

\subsection{Uncertainty as a
statistic}\label{uncertainty-as-a-statistic}

Uncertainty is also often treated as another statistic, for example, the
exceedance probability map of spatial data (see \citet{Kuhnert2018} with
software available in \citet{Lucchesi2021}), where the probability of
exceeding a specified value is displayed on the map, placing the focus
on extreme events. For risk communication, \citet{Spiegelhalter2017}
recommends showing probability with a set of coloured icons to
communicate the relative number of affected individuals. The summary
plot \citep{Potter2010} was also developed to address concerns about
uncertainty in data displays. This dense display combines the plot of a
single set of numeric values with a density estimate, boxplot, and
statistical moments.

Some authors take this approach because they explicitly believe
uncertainty is a variable of importance \citep{Blenkinsop2000}, while
others straddle the line, asserting uncertainty is acting as signal and
noise, and should fulfil both roles \citep{geointerviews}. Simply put,
these approaches might be described as swapping out the statistic for an
``uncertainty statistic'' to get an ``uncertainty visualisation''. Is it
really that easy?

\subsubsection{Example: visualising
variance}\label{example-visualising-variance}

\begin{figure}

\centering{

\includegraphics[width=1\textwidth,height=\textheight]{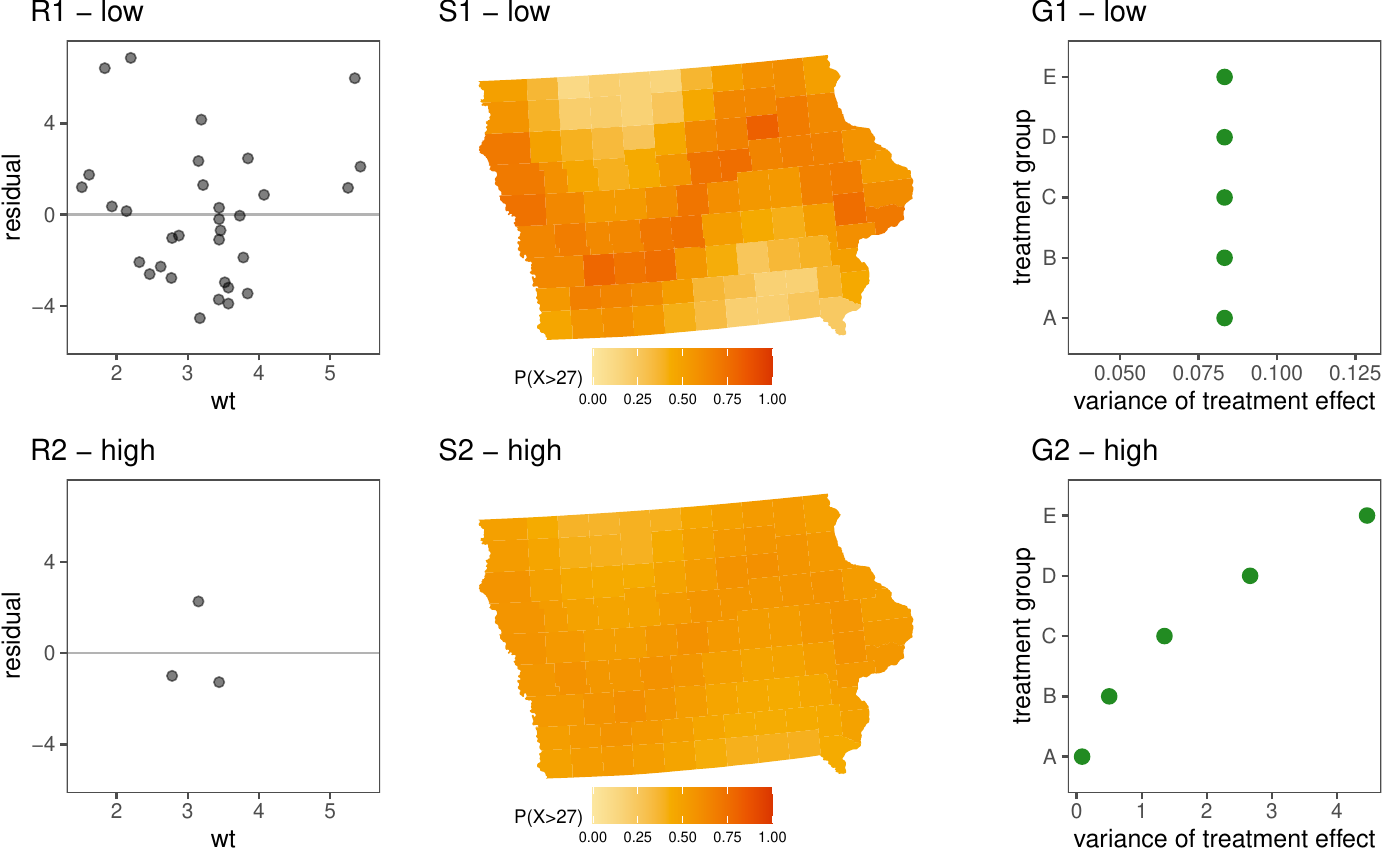}

}

\caption{\label{fig-statistic}Treating the uncertainty as a statistic,
with the same six examples. The regression (R1 and R2) is a scatterplot
of residuals vs explanatory variable, separated from the regression
making comparison of uncertainty related to the trend more difficult.
For the choropleth (S1, S2): instead of temperature, the probability of
exceeding 27\(^o\)C is shown. This has the effect of highlighting (sine
wave) trend in the low error data, and de-emphasising it in the high
error data. The plots G1 and G2 have replaced the treatment effect with
the variance on our treatment effect. It is a bit nonsensical, but we
learn something interesting that was not seen earlier: the variance in
G2 is not uniform like that in G1.}

\end{figure}

Figure~\ref{fig-statistic} depicts the six plots showing this approach
for the data introduced in Figure~\ref{fig-ignore}. The original central
value estimate has been replaced with an uncertainty statistic. Plots R1
and R2 show the residual plot of our linear regression instead of the
scatterplot and the regression line. The visual patterns we are looking
for in this plot are distinct from the linear regression, so it is hard
evaluate it relative to the trend. Sometimes the variance approach is
still related to our original display, as we can see in the exceedance
probability maps depicted in plots S1 and S2. These map
\(P(temperature>27)\) to colour for each county. The sine wave trend is
clearly visible when the error is low, but barely visible with high
error. Our visualisation of the univariate groups does reveal something
interesting: the variance is constant in G1 but different for each group
in G2. It is a nonsensical display, though, because the trend has
completely disappeared. In practice, two plots are typically presented:
one showing the main estimate and the other showing the uncertainty.

\subsubsection{What is an uncertainty visualisation,
then?}\label{what-is-an-uncertainty-visualisation-then}

What an uncertainty visualisation is or is not is one of the most
pervasive divides in the literature. For example, \citet{Leland2005}
mentions that popular graphics, such as pie charts and bar charts, omit
uncertainty. \citet{Wickham2011} suggests their product plot framework,
for area plots like bar charts and also histograms, needs to be extended
to include uncertainty representation. However, pie charts, bar charts
and histograms have all been used in a significant number of experiments
as examples of an ``uncertainty visualisation''
\citep{Ibrekk1987, Olston2002, Zhao2023, Hofmann2012}. What is going on
here?

This conflict stems from a subconscious disagreement about the purpose
of uncertainty visualisation. If you believe uncertainty visualisation
is about communicating risks or random variables, uncertainty
visualisations are just visualisations of ``uncertainty statistics'' or
distributions. On the other hand, if you believe uncertainty
visualisation is about suppressing false signals visually, then you see
an uncertainty visualisation as a transformation of an existing graphic
that adds the uncertainty in. The former has no limitation on the visual
appearance of an ``uncertainty visualisation'', allowing pie charts, bar
charts or histograms, so long as the graphic is visualising
``uncertainty'', while the latter believes uncertainty visualisations
only exist in relation to some ``normal'' visualisation. When we refer
to the graphics depicted in Figure~\ref{fig-statistic} as ``uncertainty
visualisations'', we are classifying visualisations by the data they
display, not their visual features. This is not the standard approach in
statistical graphics. A scatter plot that compares means and a scatter
plot that compares variances are both scatter plots.

Unlike plots, which are not defined by a statistic, uncertainty is
\emph{only} defined in relation to our visual statistic, a topic that
frequently appears in the literature to be dependent on the ``goals'' of
our analysis. \citet{Meng2014} commented that what is kept as data and
what is tossed away is determined by the motivation of an analysis -
what was previously noise can become signal depending on the question.
\citet{Otsuka2023} suggested that the process of observing data to
calculate statistics is largely dependent on our goals, because the
process of boiling real-world entities down into probabilities depends
on the relationships we seek to identify within our data.
\citet{Wallsten1997} argue that the best method for evaluating or
combining subjective probabilities depends on the uncertainty the
decision-maker wants to represent, and why it matters.
\citet{Fischhoff2014} suggested we should have methods for communicating
uncertainty depending on what the user is supposed to do with it.
\citet{Spiegelhalter2017} says we ``cannot assess the quality of risk
communication unless the objectives are clear''. \citet{geointerviews}
asserted that whether or not uncertainty is a source of doubt depends on
the context. The sentiment behind this repeated point is clear: the role
of uncertainty or signal is not dependent on the ``type'' of statistic,
on the source of the information, or the methods we use; it is
determined by the statistic we wish to draw inference on. Therefore, the
fundamental problem with the ``uncertainty statistic'' approach is that
the uncertainty in the plot isn't acting as noise; it is acting as
signal.

If the uncertainty in a graphic is acting as a signal, there isn't an
interesting perceptual challenge associated with the visualisation: the
uncertainty can be displayed using standard principles of graphic
design. In changing the inferential statistic, we also haven't dealt
with the original problem of integrating noise, as these ``uncertainty
statistics'' \emph{also have associated uncertainty in the estimates}
(e.g.~variance of standard deviation estimate) that is being ignored.
There is nothing wrong with explicitly visualising variance, error,
bias, or any other statistic. These metrics provide important and useful
information for analysis and decisions. The problem with this approach
is that it means everything is an uncertainty visualisation, and if
everything is an uncertainty visualisation, nothing is.

\subsection{Uncertainty as a variable}\label{uncertainty-as-a-variable}

Another common characterisation of uncertainty is as just another
variable to be integrated into the visualisation, which means
uncertainty visualisation is, at its core, a high-dimensional
visualisation problem \citep{Griethe2006, geointerviews}. This occurs
within computer science \citep{Kinkeldey2014}, cartography
\citep{MacEachren2005}, and statistical graphics \citep{Leland2005}.

Discussion of the visualisations focuses on how ``integrated'' the
uncertainty is with the estimate. \citet{Kinkeldey2014} identified a
split between intrinsic plots, where we map uncertainty to the colour or
size of the geometric object of our estimate, and extrinsic plots, where
uncertainty is mapped to a separate geometric object, such as glyphs or
error bars. Similarly, \citet{uncertchap2022} classified uncertainty
visualisations as graphical annotations (extrinsic), and probability
mapped to a visual encoding channel (intrinsic), or a hybrid of the two.
It is unclear if these levels of ``integration'' in a plot design affect
its ability to suppress signals.

\subsubsection{Example: mapping two independent
variables}\label{example-mapping-two-independent-variables}

\begin{figure}

\centering{

\includegraphics[width=1\textwidth,height=\textheight]{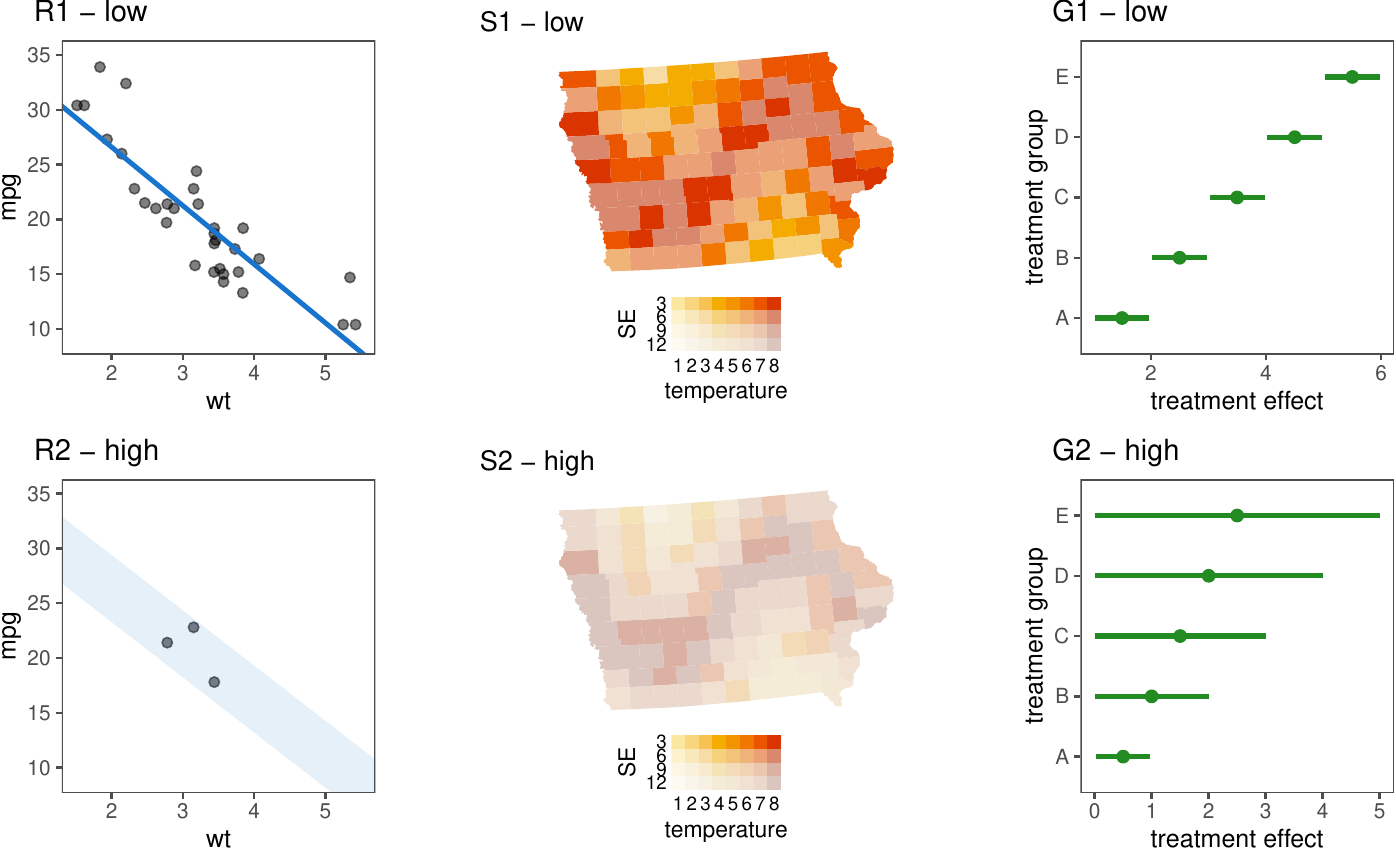}

}

\caption{\label{fig-variable}Treating the uncertainty as a variable,
with the same six examples. In plots R1 and R2, the standard error of
the slope is mapped to the line width. The standard error of the
intercept is mapped to the transparency, which is less conventional. In
plots S1 and S2, a bivariate colour palette is used with mean mapped to
the hue, and error mapped to saturation. Plots G1 and G2 represent the
mean of each group as a point, and the variance as an interval. The
uncertainty is integrated well with the signal, but for the choropleth,
the result is undesirable: the signal is easier to see when the error is
higher.}

\end{figure}

Figure~\ref{fig-variable} shows the six examples introduced in
Figure~\ref{fig-ignore}. Here, uncertainty is mapped to a spare
aesthetic in the plot. In the grammar of graphics, variables are mapped
to aesthetics, like position, colour, and size, within a plot. In plots
R1 and R2, the standard error of the slope is represented by the width
of the line. This is common, but it fails to represent the standard
error of the intercept alongside the slope. We might think we can
examine the width of the line at \texttt{wt=0}, but this would be
incorrect. Here, it has been included, less obviously, by mapping the
standard error of the intercept to transparency. While these help to
give a gestalt of the uncertainty, they are not exacting
representations. One would expect that the width above and below the
line is one standard error, but why not map the width to two standard
deviations, or even three? Interpreting transparency into a numerical
quantity is also virtually impossible, so using this aesthetic mapping
is effectively useless.

A bivariate colour palette map is shown in plots S1 and S2. With two
dimensions of the plot reserved for spatial position, it is difficult to
incorporate the error. The bivariate colour palette maps the estimate to
hue and the error to saturation, keeping the signal and noise contained
to one visual aesthetic. This has the unfortunate effect of making the
signal appear stronger when the error is higher (S2). Colour perception
is a wild beast that is hard to tame.

The extrinsic approach, shown in plots G1 and G2, has the estimate
represented by a point, and uncertainty computed as a 95\% confidence
interval mapped to line length. The trend is still visible in both
displays. It could be argued that where the variance is high, the trend
remains a main focus; that is, the display fails to sufficiently
suppress the signal.

Because all of these graphics visualise the distribution's estimate and
uncertainty as two separate pieces of information, the message of the
plot is ``here is the trend \emph{and} here is the uncertainty''. It is
worthwhile to examine why this occurs, to see if we can move towards a
version of this plot where we are able to communicate signal and noise
simultaneously.

\subsubsection{Can we visualise a ``single integrated uncertain
value''?}\label{can-we-visualise-a-single-integrated-uncertain-value}

The reality is, based on our discussion on inferential statistics,
uncertainty \emph{isn't} a separate variable: it is a component of the
random variable that is indistinguishable from the random variable
itself. Similarities between the ``as a variable'' approach and the ``as
a statistic'' approach are apparent when we read motivations for
visualising an estimate (i.e. Figure~\ref{fig-ignore}) and variance
(i.e. Figure~\ref{fig-statistic}) side-by-side using two separate
graphics. This organisation is vulnerable to change blindness
\citep{simons1997} as one needs to switch focus between two displays.)
The preference for the methods utilised in Figure~\ref{fig-variable} is
usually motivated by the difficulties in combining information on two
separate graphics \citep{moritz2017trust, Correll2018}, rather than an
understanding that the ``uncertainty statistic'' approach is not
philosophically sound. To achieve signal suppression, we need to
visualise noise and signal together as a ``single integrated uncertain
value'' \citep{Kinkeldey2014} rather than as two separate statistics.

Just because we can still see the signal in Figure~\ref{fig-variable},
that does not mean the reading of the estimate is completely independent
of the uncertainty. When making any visualisations, we usually want the
visual channels to be separable, that is, we don't want the data
represented through one visual channel to interfere with the others
\citep{Smart2019}. It is also interesting to know whether readers can
see all structures when there are multiple structures present in a data
plot, or whether they systematically fixate on one
\citep{vanderplas2017}. Separability may be desirable in standard data
visualisation, but in uncertainty visualisation, it allows the estimate
and its variance to be read independently, potentially leading to the
uncertainty being ignored \citep{uncertchap2022}. Therefore, rather than
trying to maintain visual separability, the goals of uncertainty
visualisation align far better with the pursuit of visual integration.
In an ideal system, our estimate and uncertainty would be manipulated
separately, but would be so \emph{well-integrated that they are read as
a single channel by the human brain}. The problem is that even if we can
implement the most extreme versions of integrable, our methods fall
short, as illustrated by the bivariate colour palette map in
Figure~\ref{fig-variable}. Colour hue and brightness are one of the
classic examples of integrable variables \citep{Vanderplas2020}, and
decreasing saturation should make the colours harder to distinguish, but
the signal is still clearly visible with the high variance. This is to
say nothing of the fact that multi-dimensional colour palettes can make
the graphics harder to read and less accessible \citep{Vanderplas2015}.

\subsubsection{Another example: mapping combined
variables}\label{another-example-mapping-combined-variables}

\begin{figure}

\centering{

\includegraphics[width=0.8\textwidth,height=\textheight]{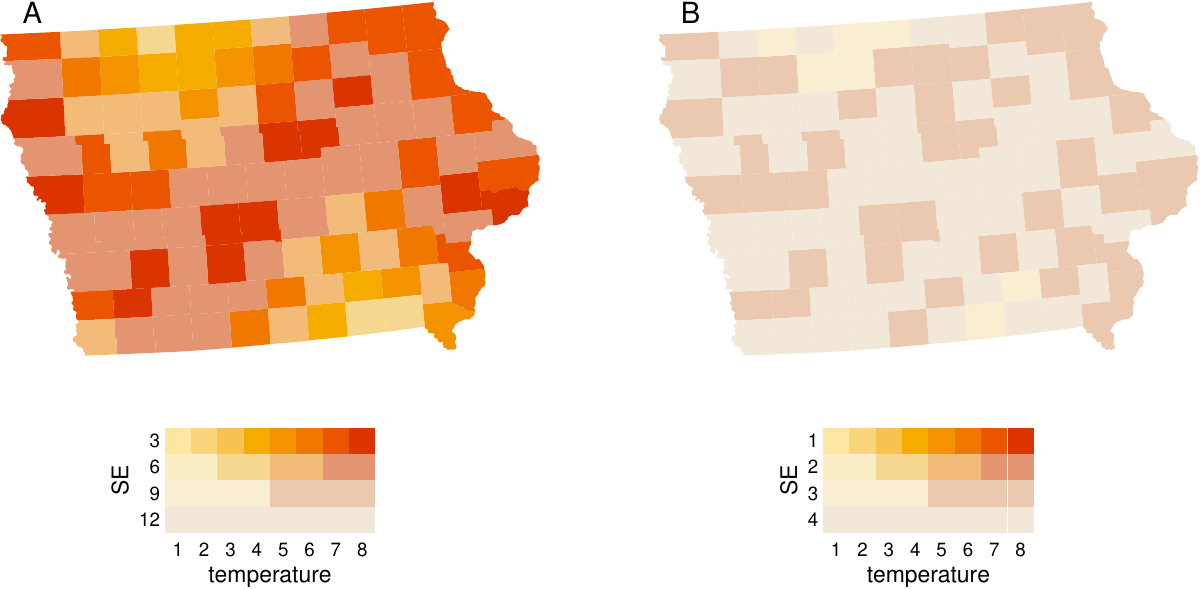}

}

\caption{\label{fig-vsup}The spatial examples displayed with a
choropleth map using a VSUP colour palette, where hue is blended when
increased uncertainty. Plot B has successfully produced signal
suppression. Look closely at the scales, though: it may have another
explanation.}

\end{figure}

The Value Suppressing Uncertainty Palette (VSUP) \citep{Correll2018} was
designed with the intention of preventing high uncertainty values from
being extracted from a map by blending colours together as they become
less certain. Figure~\ref{fig-vsup} shows the spatial example (S1, S2)
using the VSUP approach. Since the palette was designed with the
extraction of individual values in mind and it has only been tested on
simple value extraction tasks \citep{Correll2018} or search tasks
\citep{Ndlovu2023}, we can see that, at least for our example, when the
uncertainty is high, the spatial trend has functionally disappeared.

Finally, we have signal suppression! Well, not really, sorry, we tricked
you. The two plots depicted in Figure~\ref{fig-vsup} actually show the
exact same data; they are both the low variance case. If you look
closely, you can see that the two plots have a different scale, where
plot A has been scaled according to our existing knowledge about this
data, while plot B has been scaled using the range of the data passed to
the plot. Since the variance and estimate are scaled independently,
arbitrary differences in the range of our variance, unrelated to the
estimate itself, will have significant impacts on the visual appearance
of our plot. The scale issue in VSUP maps was also recognised by
\citet{Kay2019}, who noted that the suppression of any one hypothesis
largely depends on the methods we use to combine the palette, and the
variance levels at which the blending occurs. This means that, for us to
know that our plot will successfully perform signal suppression, we need
to already know what signal we are trying to suppress and set up the
VSUP palette accordingly. This means that VSUP maps are not suitable for
exploratory data analysis.

\subsubsection{Uncertainty and exploratory data
analysis}\label{uncertainty-and-exploratory-data-analysis}

The lack of uncertainty in descriptive statistics is due to the lack of
inference. Descriptive statistics are actually a small piece of a much
larger field, exploratory data analysis (EDA), that tends not to perform
statistical inference. \citet{Tukey1977} described EDA as the process of
searching for interesting hypotheses (``the greatest value of a picture
is when it forces us to notice what we never expected to see''), and
defined it in relation to confirmatory data analysis (CDA), the process
of verifying a hypothesis. There are more subfields of EDA: initial data
analysis \citep{huebner2016, chatfield}, which involves checking
assumptions and data quality prior to CDA, and model diagnostics (e.g.
\citet{bkw1980}), including posterior checks of model fit. What binds
these pursuits together is their reliance on visual summaries for making
assessments and an absence of formal inference.

\citet{Hullman2021} argued that the EDA and CDA are not entirely
distinct, as it is often difficult to draw a hard line. Our belief, as
with many concepts, is that these approaches exist on a continuum, where
we have an inherent trade-off between the number of hypotheses we can
look for and the certainty of any conclusions reached. It can help to
think of the knowledge-generating process of EDA and CDA as the nozzle
on a hose with multiple spray options, where EDA is a fine misting spray
that touches everything in the room, and CDA is a high-pressure jet
capable of obliterating any and all debris from any single spot.

Viewing EDA and CDA as a dichotomy can create some confusion when it
comes to understanding the source of uncertainty in our analysis. This
is why we have avoided the topic until now, despite the fact that an
uncertainty visualisation system for EDA is one of the most discussed
topics in the field
\citep{Hadjimichael2024, geointerviews, MacEachren2005, Sarma2024, Griethe2006}.
The EDA versus CDA dichotomy can be compared to the dichotomy between
induction, for building theories, and deduction, for testing theories,
from formal logic. One of the trade-offs in the two methods is that
deductive conclusions provide certainty, while conclusions from
induction are inherently uncertain. This means that EDA, the inductive
counterpart, has uncertainty in any conclusions reached, with the
requirement to follow up our newfound hypothesis with CDA if we want
true certainty. This adds another layer of confusion to the study of
uncertainty visualisation: conflating the uncertainty in our data with
the uncertainty that is inherent to an exploratory process (EDA).

Authors often over-compensate for the inherent EDA uncertainty,
pre-emptively hedging against every false inference that could possibly
be drawn from a graphic. \citet{Hullman2021} argues there is no such
thing as a ``model-free'' visualisation. \citet{Guo2024} provides a
grammar for visualising statistical model checks. The lineup protocol
\citep{Buja2009} provides the viewer with plots of the data in a field
of plots of null data where any patterns seen are due to sampling
variability. The Rorschach protocol, from the same paper, shows only
null plots to give the reader some intuition for what spurious sampling
patterns exist. \citet{Savvides2019} provides a statistical super-test
against multiple comparisons driven by probabilistic arguments. There is
a CDA quality to these approaches. The sum total of a lot of CDA is not
EDA, just as swinging a high-pressure jet around a room is not
equivalent to using a misting spray. While EDA and CDA may be along a
continuum, we cannot simultaneously perform EDA and CDA as the
approaches are, philosophically speaking, perpendicular to one another.

To truly create an uncertainty visualisation approach that is capable of
EDA, we need to accept that uncertainty is inherent to the method, and
it cannot be pre-emptively removed from the visualisations. If we accept
this fact, then the only possible source of uncertainty in an
uncertainty visualisation system that performs EDA is from the data
itself. That is to say, the only possible way for there to be
uncertainty in a visualisation designed for EDA is that the data itself
represents inference that has been done earlier in our analysis. We can
see this in the original description of Figure~\ref{fig-ignore}, where
the uncertainty in all cases, even measurement error, represents
inference that was performed earlier in our analysis.

In the VSUP approach, our ability to arbitrarily decide which values to
blend at, or which suppression approach to use, means that the
``uncertainty'' we are visualising will be informed by the conclusions
we are drawing, and not a product of the data itself -- an antithetical
approach to EDA. For a visualisation to be suitable for EDA, it should
always look the same regardless of what hypothesis we plan to draw from
it. As much of the transparency in data visualisation comes from this
feature in EDA, it is reasonable to set it as a requirement of our
visualisations. Ensuring this property means we cannot treat signal and
noise as separate variables, but rather as a single integrated unit.

\subsection{Uncertainty as a
distribution}\label{uncertainty-as-a-distribution}

Rather than trying to reduce random variables to a single value or pair
of values, why not visualise the whole distribution? This approach is
found in computer science's hypothetical outcome plots (HOPs), which
animate a sequence of potential outcomes of a distribution
\citep{Hullman2015}, in geoscience's pixel-maps
\citep{Lucchesi2021, Blenkinsop2000}, and in statistics multiple
forecasts \citep{fpp3}.

\subsubsection{Example: visualising
samples}\label{example-visualising-samples}

Figure~\ref{fig-variable} shows the six examples introduced in
Figure~\ref{fig-ignore}, shown as distributions. The distribution is
represented by either a sample of outcomes or a quantile dot plot. Plots
R1 and R2 show a linear regression as a sample of possible outcomes from
the distribution. The distribution used for both the slope and intercept
is normal with the conventional mean and standard error. Plots S1 and S2
show a pixel map, which is a choropleth map where each county is
coloured by a sample of outcomes from the distribution. Plots G1 and G2
show each univariate distribution as a quantile dot plot. We can see
that the strong downward trend in the linear regression, the sine wave
in the choropleth map, and the incremental increase in the univariate
distributions are all clearly visible in the low variance case, but
disappear in the high variance cases. The graphics have achieved signal
suppression. Visualising the random variable as a distribution gives
additional information, such as the previously hidden bimodality of the
univariate distributions in G1 and G2.

\begin{figure}

\centering{

\includegraphics[width=1\textwidth,height=\textheight]{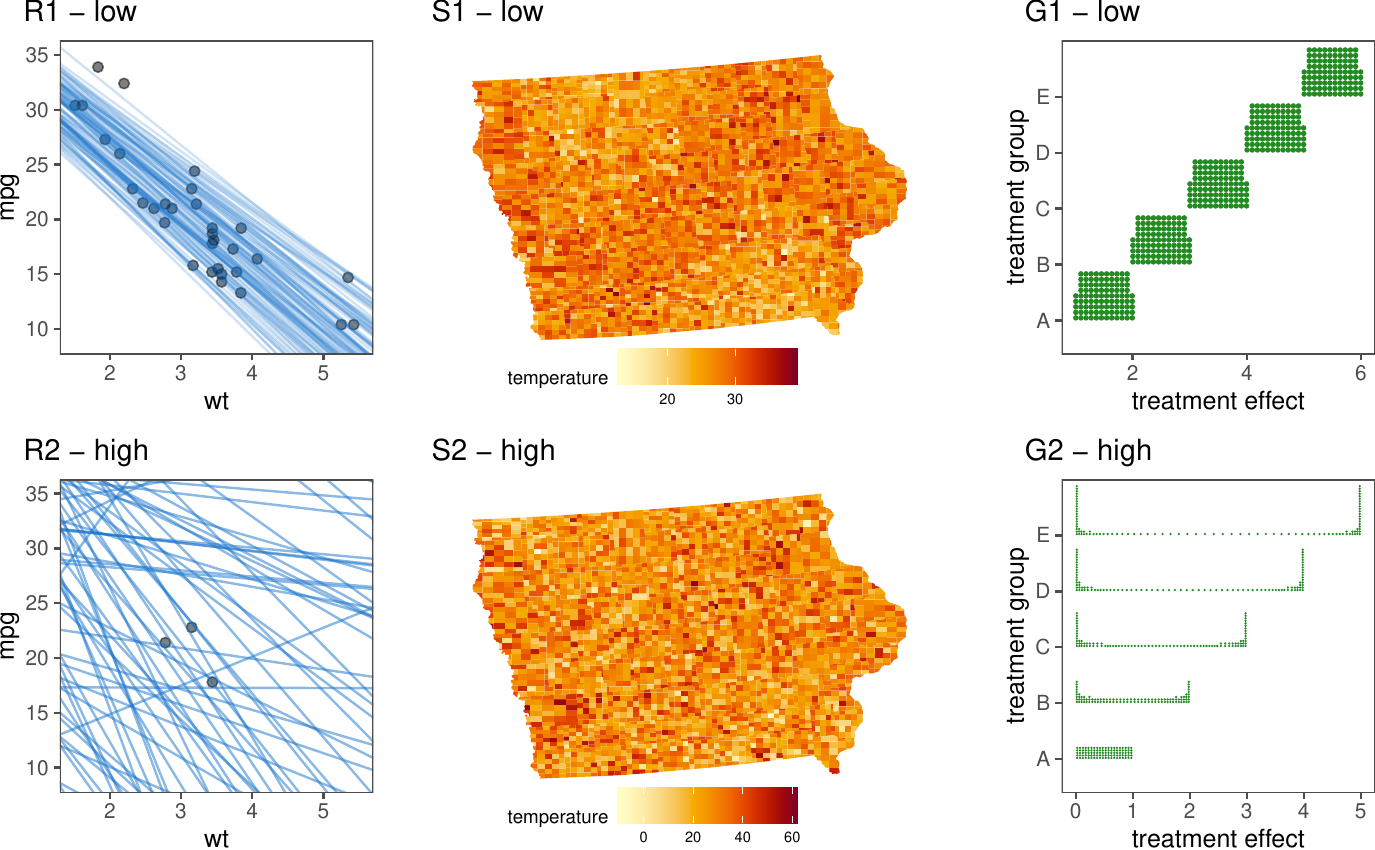}

}

\caption{\label{fig-distribution}Treating the uncertainty as a
distribution, with the same six examples. Plots R1 and R2 show a linear
regression as a sample of possible outcomes from the distribution. Plots
S1 and S2 show a pixel map. Plots G1 and G2 show the groups as quantile
dot plots. In each case, the signal (regression line, sine wave,
increasing trend) has disappeared with high uncertainty.}

\end{figure}%

\subsubsection{Quantified versus unquantified
uncertainty}\label{quantified-versus-unquantified-uncertainty}

By showing our distribution as ``data'', we are able to read the
``uncertainty'' plots using the same perceptual mechanisms we use to
read the ``non-uncertainty'' plot. This should lead to more effective
communication, as people tend to read more complicated visualisations
like the bivariate and VSUP plots the same way they read the simple
choropleth counterpart \citep{Ndlovu2023}. This approach also does not
significantly hinder our ability to extract the individual values mapped
by the previous plots, as extracting global statistics from a sample can
be done with relative ease \citep{Franconeri2021}. We are able to
include more information by offloading more computation to visual
processing, but what is the limitation on this approach? Which aspects
of uncertainty should be processed by our visual system, and which
should be processed by our statistical computation? It is not obvious
from the question, but this is actually a question about how much of our
uncertainty should be quantified.

Quantified uncertainty usually focuses narrowly on concepts such as
probability, confidence intervals, variance, error, or precision
\citep{Hullman2018, Maceachren2012, Thomson2005}, while unquantified
uncertainty often includes a broader range of concepts like missing
values, reliability, model validity, or source integrity
\citep{Griethe2006, Leland2005, Pang1997, Pham2009, Boukhelifa2017}.
When discussing uncertainty, we typically include these unquantified
uncertainties, not because these things \emph{are} uncertainty, but
because they can \emph{create} uncertainty when we perform inference.
This is often because these unquantified uncertainties violate our
assumptions of the uniformity of nature \citep{Otsuka2023}.

Sometimes we are able to visualise these assumption violations directly.
For example, we can check for structure in our missing data using the
\texttt{naniar} package \citet{Tierney2023} that allows us to include
missing values as a ``shadow'' alongside our usual visualisations. This
approach amounts to just ``showing the data'', which is a simple but
effective option for uncertainty visualisation that is largely
overlooked. While this approach is useful for better understanding data,
it will not eliminate trends that have become invalid due to structure
in missing data or an invalid model. We can only integrate uncertainty
as noise when that uncertainty has been \emph{quantified} as an effect
on the estimates we are visualising. This is not to say one method is
preferable; visualising both quantified and unquantified uncertainty is
necessary for a healthy analysis. Data analysis often works in cycles,
where we find assumption violations using EDA, quantify the effect of
these violations on inference, and then visualise the output of that
inference using uncertainty visualisation.

\section{Evaluating uncertainty
visualisations}\label{evaluating-uncertainty-visualisations}

Unfortunately, the conflicting results in the field are not limited to
plot design and extend to the experimental findings as well
\citep{MacEachren2005, Kinkeldey2014, Hullman2016}. There are as many
explanations for the noisy evaluation studies as there are
contradictions in the research itself. \citet{Kim2019} believes there is
some interference in results from participants' prior beliefs;
\citet{Hullman2016} believes the noise in the literature could come from
visual heuristics, subjective probabilities, unknown participant utility
functions, or a misunderstanding of statistical concepts (such as
confidence intervals). \citet{Kinkeldey2014} suggest the perception of
visualisation changes by audience, so we cannot expect the same results
between different subpopulations. \citet{Brennen2018} attributes
evaluation difficulties to cognitive load from complicated uncertainty
visualisations, as well as the participants' prior experience in the
topic. While these issues will certainly have some impact on our ability
to synthesise, none of them is unique to uncertainty visualisation.
Rather, the issue is likely due to a disconnect between the evaluation
methods used and the stated goals of each experiment, a common issue in
visualisation evaluation experiments \citep{Vanderplas2020}.

\subsection{Current evaluation
methods}\label{current-evaluation-methods}

\subsubsection{Value extraction}\label{value-extraction}

Uncertainty visualisations are most commonly evaluated based on how
accurately viewers can extract an estimate and its variance
\citep{Kinkeldey2014, Hullman2019}. This is not unusual, as direct
observation is the simplest way to verify that information can be
accurately read from a graph \citep{Vanderplas2020}. Unfortunately, this
approach doesn't work for uncertainty visualisation. The second we ask a
specific question about a statistic, that statistic becomes inferential,
even if the plot was not the intent behind the question. By shifting the
focus from \(\hat{X}\) to \(Var(\hat{X})\) or \(P(\hat{X}<x)\), we end
up evaluating visualisations on their ability to convey uncertainty
statistics, rather than their ability to perform signal suppression.
Even if the authors do not realise it themselves, there is nothing
unique to uncertainty in these studies, so when we boil the findings
down to generalised results, they simply restate existing principles
within information visualisation. Some of the findings are obvious:
participants were more accurate when reading a probability expressed as
text than when they had to extract it from a graphic
\citep{Cheong2016, Savelli2013}. Other studies replicate existing
research, such as the finding that a probability mapped to a position is
more accurate than one mapped to an area
\citep{Ibrekk1987, Gschwandtnei2016}, established as part of the
hierarchy of perceptual tasks more than 40 years ago
\citetext{\citealp{Cleveland1984}; \citealp[replicated by][]{heer2010}}.
This extends beyond simple accuracy evaluations: \citet{Sanyal2009}
found that colour was more effective than size when searching for
extrema in variances; we have known that pre-attentive aesthetics, such
as colour, are more efficient for search tasks since the 1980s
\citep{Vanderplas2020}. By classifying these studies as evaluations of
``uncertainty visualisation'' while evaluating uncertainty as a signal,
we are encouraged to see successful examples of signal suppression as
failure. This approach leads authors to advise against particular
aesthetic mappings for uncertainty, because they cause participants to
have more difficulty extracting values \citep{Blenkinsop2000}. This
conclusion is antithetical to the goals of signal suppression and occurs
because these methods evaluate uncertainty as a signal, not as noise.

\subsubsection{Trust, confidence, and risk
aversion}\label{trust-confidence-and-risk-aversion}

If we cannot directly measure uncertainty for fear that it turns into a
signal, we might then assume we can measure the secondary benefits of
increased transparency. This seems to be the approach of many
visualisation authors, as secondary benefits such as trust, confidence,
and risk aversion are all frequently used in uncertainty evaluation
studies \citep{Hullman2019}. Unfortunately, measuring these secondary
effects often leads to confusing conclusions that simultaneously argue
for and against the inclusion of uncertainty.

This is most commonly noticed in the use of trust as a measure, as
several authors have commented that measuring trust, and not
transparency, can lead to a questionable subtext that argues against
transparency \citep{Spiegelhalter2017, ONeill2018}. We see this directly
play out in the visualisation literature, where surveyed visualisation
authors explicitly said they didn't include uncertainty due to the fact
that they might decrease trust in their conclusions
\citep{Hullman2020a}. This sentiment is also true for confidence, as
\citet{Blenkinsop2000} commented that visually integrable depictions of
uncertainty should be avoided, as they decrease the viewer's confidence
in their extracted data values.

Another secondary effect that is similar to trust and confidence is risk
aversion. Risk aversion is an economic term used to describe an agent
who would choose a random variable with a lower expected payout because
it also has a lower variance. Risk aversion's role in the uncertainty
visualisation is unclear, as authors will argue uncertainty should
elicit more risk aversion in one paper \citep{Hullman2019}, and argue
for less risk aversion (by proxy of suggesting rational agents as a
benchmark) in the next \citep{wu2023rational}. Ultimately, these
approaches have similar issues to value extraction studies, except they
are slightly more confusing in their goals, leading them to
simultaneously argue for and against the inclusion of uncertainty in a
visualisation.

\subsubsection{Alternative approaches}\label{alternative-approaches}

Often, authors understand that the effects of uncertainty are more
complicated than simple value extraction. These studies indicate that
accurately capturing uncertainty will be more complicated than simply
avoiding value extraction or trust as a measure.

One approach is to ask indeterminate questions, such as asking
participants for the ``best estimate'' \citep{Ibrekk1987}, or to select
which distribution is the ``furthest to the right'' in a lineup
\citep{Hofmann2012}. In both cases, the ground truth is based on the
mean of the distribution, which is not as indeterminate as the question.
This approach can lead to inconclusive results, as we are left unclear
whether it was the phrasing of the question or the plot design that
caused the participants to answer incorrectly.

On the other hand, questions that are incredibly specific about the
distribution information can confuse the participants and induce noisy
results. For example, \citet{Hullman2015} asked participants to compare
two normally distributed groups, A and B, and had many participants say
that group A was more likely to be bigger, despite group B having a
higher mean. \citet{Gschwandtnei2016} asked participants the
``probability that the interval has already ended at the marked point in
time?'' and participants replied with the probability that the interval
had already started.

The confusion around trying to capture the effects of uncertainty can
also (understandably) extend to the authors of the study itself. We can
see an example of this in \citet{Padilla2017}. In order to answer the
question correctly, the first experiment required participants to assume
that an oil rig being ``more likely to be hit'' by a hurricane would
\emph{not} translate to the rig sustaining ``more damage''. The second
experiment required participants to assume the opposite.

\subsubsection{Effective methods}\label{effective-methods}

This is not to say all evaluation studies fail to properly evaluate
uncertainty as noise. There are several studies that ask participants to
identify a particular signal that the noise is trying to obfuscate
\citep{Kale2018, Correll2014}, which seems to be an effective method.
The only problem with these studies is that most uncertainty
visualisation methods are not integrated into \emph{the grammar of
graphics} \citep{Leland2005, ggplot2}, so we regularly see comparisons
between disparate plots that would never be considered substitutes for
one another outside the artificial ``uncertainty visualisation''
framework they are placed within. For example, \citet{Kale2018} compared
static bar charts with error bars to a bar chart with animated samples,
meaning that any difference in participants' ability to read the plot
could be due to the statistic (confidence interval versus sample), the
geometry (bars versus intervals), or the use of animation (static versus
animated plots). This issue was rectified in their second experiment,
where they compared overlayed and animated samples, giving us an insight
into the types of visualisations that are appropriate to compare in
uncertainty visualisation experiments. This means that even when
evaluations \emph{are} done correctly, there is no generalisable theory
we can take from the results.

The point here is not to accuse the authors of poor academic rigour. The
papers are (usually) logically consistent and well-formulated pieces of
work. Rather, the point is to illustrate that evaluating uncertainty as
noise is surprisingly difficult. Designing tests for signal suppression
will require a formalisation of uncertainty within the grammar of
graphics, as well as improved evaluation methods.

\subsection{Implicit Hypothesis
Testing}\label{implicit-hypothesis-testing}

The main problem with current uncertainty visualisation evaluations is
that they often require explicit (or convoluted) questions about the
variance. Asking direct questions about the statistics or outcomes is
not an explicit requirement of visualisation evaluations. In their
review of testing statistical graphics, \citet{Vanderplas2020} drew a
distinction between explicit tests, where participants are asked direct
questions about specific features of a plot, and implicit testing, where
users identify both the purpose and function of the plot. The lineup
protocol is the most salient example of the implicit approach. Lineups
are a confirmatory visualisation tool where participants are shown a set
of \(M\) plots, and asked to identify the plot that is the ``most
different'', leaving participants to decide what ``most different''
means to them, even if it is not what the authors intended
\citep{vanderplas2017}. The implicit test does not limit the versatility
of the approach, with the lineup being used to evaluate the
effectiveness of different types of plots \citep{Hofmann2012}, colour
palettes \citep{Reda2021}, and design decisions \citep{vanderplas2017}.

Lineup protocols are not only useful for implicit testing: they also
have parallels to hypothesis testing that can be leveraged in
uncertainty visualisation. The concept of signal suppression is, at its
core, an assertion of statistical validity: the visibility of signals
should be directly proportional to \(p\)-values or some equivalent
measure. This comparison is not new in uncertainty visualisation, where
parallels have been drawn to frequentist statistics by
\citet{Correll2014}, who compared results to Cohen's D, and to Bayesian
statistics by \citet{Kim2019}, who evaluated plots based on their impact
on the users' prior beliefs. The comparison to hypothesis testing is far
more natural for the lineup protocol, which has a visual test statistic
\citep{Majumder2013} and can be compared to standard statistical tests
using power curves \citep{Patrick2023}. The connections between lineups
and uncertainty visualisation are numerous and have been previously
identified in the development of HOPs \citep{Hullman2015}.

The lineup protocol and uncertainty visualisations are similar: lineups
were designed for checking if perceived patterns are real or merely the
result of chance \citep{Buja2009, Wickham2010}. As both approaches are
attempting to do the same thing, it is likely that we are unable to
leverage the lineup protocol directly to evaluate uncertainty
visualisation, but a new evaluation methodology should try to learn from
the success of the lineup approach. Designing an implicit testing method
for uncertainty visualisation that allows us to draw parallels to
standard notions of statistical significance would solve many of the
issues with the current evaluation approaches.

\section{Conclusions and Future Work}\label{conclusions-and-future-work}

This paper examines the literature and provides suggestions for a
structural framework to support uncertainty visualisation. Particularly,
we propose that uncertainty visualisation should accomplish signal
suppression, dampening weak signals and amplifying strong signals. We
have also highlighted several gaps in the existing literature.

\emph{Experimental practices on uncertainty visualisation need
standards.} Some existing evaluation experiments treat uncertainty as a
signal, while others treat uncertainty as noise. As a result, it is
difficult to combine results from papers to get a meaningful sense of
how uncertainty information is understood by a viewer. Researchers need
to ensure that when they identify the motivation behind their
visualisation technique, their evaluation methods align with the stated
goals of the paper.

\emph{Experimental methods that evaluate uncertainty as noise need to be
developed.} Research into separability and integrability of signal and
noise is of particular interest to uncertainty visualisation, as it
allows assessment of the interference between the two. When designing
experiments, authors often choose aesthetics that are visually
distinguishable; uncertainty visualisation authors should consider doing
the opposite.

\emph{Uncertainty needs to be formalised within the grammar of
graphics.} Some of this formalisation was done by \citet{Kay2023}, but
it focuses only on the visualisation of univariate distributions. Giving
authors the ability to describe uncertainty visualisations in terms of
statistics, geometries and aesthetics will support evaluation
experiments that can build towards a cohesive theory of visualising
uncertainty.

\emph{Software that allows users to easily perform signal suppression is
needed.} Existing uncertainty visualisation methods view a distribution
as its own object, and there are no software options treating ``an
uncertainty visualisation as a function of an existing visualisation''
philosophy.

Signal suppression is an undeveloped area of visualisation research, and
developing methods for the practice may require us to challenge our
entire notion of what makes a good visualisation.

\section*{Reproducibility}\label{reproducibility}
\addcontentsline{toc}{section}{Reproducibility}

The R packages were used for this work were: \texttt{tidyverse}
\citep{tidyverse}, \texttt{RColorBrewer} \citep{RColorBrewer},
\texttt{scales} \citep{scales}, \texttt{sf} \citep{sf},
\texttt{urbnmapr} \citep{urbnmapr}, \texttt{flextable}
\citep{flextable}, \texttt{colorspace} \citep{colorspace},
\texttt{ggdist} \citep{Kay2023}, \texttt{ggdibbler} \citep{ggdibbler},
\texttt{patchwork} \citep{patchwork}, \texttt{distributional}
\citep{distributional}, \texttt{ggthemes} \citep{ggthemes},
\texttt{broom} \citep{broom}, and \texttt{rgeos} \citep{rgeos}. The
GitHub repository for this paper can be found at
https://github.com/harriet-mason/ARSA-UncertaintyLitReview, which
contains the files required to reproduce this article in full.

\renewcommand{\bibsection}{}
\bibliography{paper.bib}

@article{Meng2014,
  author = {Meng, Xiao Li},
  doi = {10.1201/b16720-52},
  isbn = {9781482204988},
  journal = {Past, Present, and Future of Statistical Science},
  pages = {537--562},
  title = {A trio of inference problems that could win you a nobel prize in statistics (if you help fund it)},
  year = {2014}
}

@article{utypo,
  author = {Walker, W. E. and Harremoes, P. and Rotmans, J and {Van Der Sluijs}, J. P. and {Van Asselt}, M. B. A. and Janssen, P and {Krayer Von Krauss}, M. P.},
  issn = {1389-5176},
  journal = {Integrated Assessment},
  number = {1},
  pages = {5--17},
  title = {Defining Uncertainty},
  url = {https://www.narcis.nl/publication/RecordID/oai:tudelft.nl:uuid:fdc0105c-e601-402a-8f16-ca97e9963592},
  volume = {4},
  year = {2003}
}

@article{anscombe,
	author = { F. J. Anscombe },
	title = {Graphs in Statistical Analysis},
	journal = {The American Statistician},
	volume = {27},
	number = {1},
	pages = {17-21},
	year  = {1973},
	publisher = {Taylor \& Francis},
	URL = {https://www.tandfonline.com/doi/abs/10.1080/00031305.1973.10478966}
}

@Manual{datasaurpkg,
    title = {{datasauRus}: {D}atasets from the Datasaurus Dozen},
    author = {Steph Locke and Lucy {D'Agostino McGowan}},
    year = {2018},
    note = {R package version 0.1.4},
    url = {https://CRAN.R-project.org/package=datasauRus},
}

@article{Olston2002,
  author = {Olston, C. and Mackinlay, J. D.},
  doi = {10.1109/INFVIS.2002.1173145},
  isbn = {076951751X},
  issn = {1522404X},
  journal = {Proceedings - IEEE Symposium on Information Visualization, INFO VIS},
  pages = {37--40},
  publisher = {IEEE},
  title = {Visualizing data with bounded uncertainty},
  volume = {2002-Janua},
  year = {2002}
}

@book{Otsuka2023,
  address = {New York},
  author = {Otsuka, Jun},
  doi = {10.4324/9781003319061},
  edition = {1st},
  isbn = {9781003319061},
  pages = {204},
  publisher = {Routledge},
  title = {Thinking About Statistics: The Philosophical Foundations},
  year = {2023}
}

@article{Kinkeldey2014,
  author = {Kinkeldey, Christoph and MacEachren, Alan M. and Schiewe, Jochen},
  doi = {10.1179/1743277414Y.0000000099},
  issn = {17432774},
  journal = {Cartographic Journal},
  number = {4},
  pages = {372--386},
  title = {How to assess visual communication of uncertainty? {A} systematic review of geospatial uncertainty visualisation user studies},
  volume = {51},
  year = {2014}
}

@article{Wallsten1997,
  author = {Wallsten, Thomas S. and Budescu, David V. and Erev, Ido and Diederich, Adele},
  doi = {10.1002/(sici)1099-0771(199709)10:3<243::aid-bdm268>3.0.co;2-m},
  issn = {08943257},
  journal = {Journal of Behavioral Decision Making},
  number = {3},
  pages = {243--268},
  title = {Evaluating and combining subjective probability estimates},
  volume = {10},
  year = {1997}
}

@article{Fischhoff2014,
  author = {Fischhoff, Baruch and Davis, Alex L.},
  doi = {10.1073/pnas.1317504111},
  issn = {10916490},
  journal = {Proceedings of the National Academy of Sciences of the United States of America},
  pages = {13664--13671},
  pmid = {25225390},
  title = {Communicating scientific uncertainty},
  volume = {111},
  year = {2014}
}

@article{Hullman2016,
  author = {Hullman, Jessica},
  doi = {10.1145/2993901.2993919},
  isbn = {9781450348188},
  journal = {ACM International Conference Proceeding Series},
  pages = {143--151},
  title = {Why evaluating uncertainty visualization is error prone},
  volume = {24-October},
  year = {2016}
}

@article{Spiegelhalter2017,
  author = {Spiegelhalter, David},
  doi = {10.1146/annurev-statistics-010814-020148},
  issn = {2326831X},
  journal = {Annual Review of Statistics and Its Application},
  pages = {31--60},
  title = {Risk and uncertainty communication},
  volume = {4},
  year = {2017}
}

@article{Ibrekk1987,
  author = {Ibrekk, Harald and Morgan, M. Granger},
  doi = {10.1111/j.1539-6924.1987.tb00488.x},
  issn = {15396924},
  journal = {Risk Analysis},
  number = {4},
  pages = {519--529},
  title = {Graphical Communication of Uncertain Quantities to Nontechnical People},
  volume = {7},
  year = {1987}
}

@article{Sanyal2009,
  author = {Sanyal, Jibonananda and Zhang, Song and Bhattacharya, Gargi and Amburn, Phil and Moorhead, Robert J.},
  doi = {10.1109/TVCG.2009.114},
  issn = {10772626},
  journal = {IEEE Transactions on Visualization and Computer Graphics},
  number = {6},
  pages = {1209--1218},
  pmid = {19834191},
  publisher = {IEEE},
  title = {A user study to compare four uncertainty visualization methods for {1D} and {2D} datasets},
  volume = {15},
  year = {2009}
}

@article{Gustafson2019,
  author = {Gustafson, Abel and Rice, Ronald E.},
  doi = {10.1177/1075547019870811},
  issn = {15528545},
  journal = {Science Communication},
  number = {6},
  pages = {679--706},
  title = {The Effects of Uncertainty Frames in Three Science Communication Topics},
  volume = {41},
  year = {2019}
}

@article{Thomson2005,
  author = {Thomson, Judi and Hetzler, Elizabeth and MacEachren, Alan and Gahegan, Mark and Pavel, Misha},
  doi = {10.1117/12.587254},
  issn = {0277786X},
  journal = {Visualization and Data Analysis 2005},
  number = {March 2005},
  pages = {146},
  title = {A typology for visualizing uncertainty},
  volume = {5669},
  year = {2005}
}

@article{Pang1997,
  author = {Pang, Alex T. and Wittenbrink, Craig M. and Lodha, Suresh K.},
  doi = {10.1007/s003710050111},
  issn = {01782789},
  journal = {Visual Computer},
  number = {8},
  pages = {370--390},
  title = {Approaches to uncertainty visualization},
  volume = {13},
  year = {1997}
}

@incollection{Pham2009,
  author = {Pham, Binh and Streit, Alex and Brown, Ross},
  booktitle = {Advanced Information and Knowledge Processing},
  doi = {10.1007/978-1-84800-269-2_2},
  issn = {21978441},
  pages = {19--48},
  publisher = {Springer-Verlag London Ltd},
  title = {Visualization of information uncertainty: Progress and challenges},
  volume = {36},
  year = {2009}
}

@inproceedings{Boukhelifa2017, 
  author = {Boukhelifa, Nadia and Perrin, Marc-Emmanuel and Huron, Samuel and Eagan, James}, 
  title = {How Data Workers Cope with Uncertainty: A Task Characterisation Study}, 
  year = {2017}, 
  isbn = {9781450346559}, 
  publisher = {Association for Computing Machinery}, 
  address = {New York, NY, USA}, 
  url = {https://doi.org/10.1145/3025453.3025738}, 
  doi = {10.1145/3025453.3025738}, 
  booktitle = {Proceedings of the 2017 CHI Conference on Human Factors in Computing Systems}, 
  pages = {3645–3656}, 
  numpages = {12}, 
  location = {Denver, Colorado, USA}, 
  series = {CHI '17}
}

@article{Hullman2020a,
  author = {Hullman, Jessica},
  doi = {10.1109/TVCG.2019.2934287},
  eprint = {1908.01697},
  issn = {19410506},
  journal = {IEEE Transactions on Visualization and Computer Graphics},
  month = {jan},
  number = {1},
  pages = {130--139},
  pmid = {31425093},
  publisher = {IEEE Computer Society},
  title = {Why Authors Don't Visualize Uncertainty},
  volume = {26},
  year = {2020}
}

@article{Manski2020,
  author = {Manski, Charles F.},
  doi = {10.1017/S0266267119000105},
  issn = {14740028},
  journal = {Economics and Philosophy},
  number = {2},
  pages = {216--245},
  title = {The lure of incredible certitude},
  volume = {36},
  year = {2020}
}

@article{ONeill2018,
  author = {O'Neill, Onora},
  doi = {10.1080/09672559.2018.1454637},
  issn = {09672559},
  journal = {International Journal of Philosophical Studies},
  number = {2},
  pages = {293--300},
  publisher = {Routledge},
  title = {Linking Trust to Trustworthiness},
  url = {https://doi.org/10.1080/09672559.2018.1454637},
  volume = {26},
  year = {2018}
}

@article{Zhao2023,
  author = {Zhao, Jieqiong and Wang, Yixuan and Mancenido, Michelle V. and Chiou, Erin K. and Maciejewski, Ross},
  doi = {10.1109/TVCG.2023.3251950},
  issn = {19410506},
  journal = {IEEE Transactions on Visualization and Computer Graphics},
  number = {7},
  pages = {4093--4107},
  publisher = {IEEE},
  title = {Evaluating the Impact of Uncertainty Visualization on Model Reliance},
  volume = {30},
  year = {2023}
}

@inproceedings{Griethe2006,
  title = {The Visualization of Uncertain Data: Methods and Problems},
  author = {Griethe, Henning and Schumann, Heidrun},
  booktitle={SimVis},
  volume={6},
  pages={143--156},
  year={2006}
}

@book{Leland2005,
  author = {Wilkinson, Leland},
  title = {The Grammar of Graphics},
  year = {2005},
  isbn = {0387245448},
  publisher = {Springer-Verlag},
  address = {Berlin, Heidelberg}
}

@article{Cleveland1984,
  author = {Cleveland, William S. and McGill, Robert},
  doi = {10.1080/01621459.1984.10478080},
  issn = {1537274X},
  journal = {Journal of the American Statistical Association},
  number = {387},
  pages = {531--554},
  title = {Graphical perception: Theory, experimentation, and application to the development of graphical methods},
  volume = {79},
  year = {1984}
  }

@article{Wickham2011,
  author = {Wickham, Hadley and Hofmann, Heike},
  doi = {10.1109/TVCG.2011.227},
  issn = {10772626},
  journal = {IEEE Transactions on Visualization and Computer Graphics},
  number = {12},
  pages = {2223--2230},
  title = {Product plots},
  volume = {17},
  year = {2011}
}

@article{Correll2018,
  author = {Correll, Michael and Moritz, Dominik and Heer, Jeffrey},
  doi = {10.1145/3173574.3174216},
  isbn = {9781450356206},
  journal = {Conference on Human Factors in Computing Systems - Proceedings},
  pages = {1--11},
  title = {Value-suppressing uncertainty palettes},
  volume = {2018-April},
  year = {2018}
}

@article{Buja2009,
  author = {Buja, Andreas and Cook, Dianne and Hofmann, Heike and Lawrence, Michael and Lee, Eun Kyung and Swayne, Deborah F. and Wickham, Hadley},
  doi = {10.1098/rsta.2009.0120},
  issn = {1364503X},
  journal = {Philosophical Transactions of the Royal Society A: Mathematical, Physical and Engineering Sciences},
  month = {nov},
  number = {1906},
  pages = {4361--4383},
  pmid = {19805449},
  publisher = {Royal Society},
  title = {Statistical inference for exploratory data analysis and model diagnostics},
  volume = {367},
  year = {2009}
}

@article{Wickham2010,
  author = {Wickham, Hadley and Cook, Dianne and Hofmann, Heike and Buja, Andreas},
  doi = {10.1109/TVCG.2010.161},
  issn = {10772626},
  journal = {IEEE Transactions on Visualization and Computer Graphics},
  pages = {973--979},
  pmid = {20975134},
  title = {Graphical inference for infovis},
  volume = {16},
  year = {2010}
}

@article{Hullman2018,
  author = {Hullman, Jessica and Kay, Matthew and Kim, Yea Seul and Shrestha, Samana},
  doi = {10.1109/TVCG.2017.2743898},
  issn = {10772626},
  journal = {IEEE Transactions on Visualization and Computer Graphics},
  number = {1},
  pages = {446--456},
  pmid = {28866501},
  publisher = {IEEE},
  title = {Imagining Replications: Graphical Prediction Discrete Visualizations Improve Recall Estimation of Effect Uncertainty},
  volume = {24},
  year = {2018}
}

@article{Boukhelifa2012,
  author = {Boukhelifa, Nadia and Bezerianos, Anastasia and Isenberg, Tobias and Fekete, Jean Daniel},
  doi = {10.1109/TVCG.2012.220},
  issn = {10772626},
  journal = {IEEE Transactions on Visualization and Computer Graphics},
  number = {12},
  pages = {2769--2778},
  publisher = {IEEE},
  title = {Evaluating sketchiness as a visual variable for the depiction of qualitative uncertainty},
  volume = {18},
  year = {2012}
}

@article{Potter2010,
  author = {Potter, K. and Kniss, J. and Riesenfeld, R. and Johnson, C. R.},
  doi = {10.1111/j.1467-8659.2009.01677.x},
  issn = {14678659},
  journal = {Computer Graphics Forum},
  number = {3},
  pages = {823--832},
  title = {Visualizing summary statistics and uncertainty},
  volume = {29},
  year = {2010}
}

@article{Gschwandtnei2016,
  author = {Gschwandtner, Theresia and B{\"{o}}gl, Markus and Federico, Paolo and Miksch, Silvia},
  doi = {10.1109/TVCG.2015.2467752},
  issn = {10772626},
  journal = {IEEE Transactions on Visualization and Computer Graphics},
  month = {jan},
  number = {1},
  pages = {539--548},
  pmid = {26529717},
  publisher = {IEEE Computer Society},
  title = {Visual Encodings of Temporal Uncertainty: A Comparative User Study},
  volume = {22},
  year = {2016}
}

@article{Hullman2015,
  author = {Hullman, Jessica and Resnick, Paul and Adar, Eytan},
  doi = {10.1371/journal.pone.0142444},
  issn = {19326203},
  journal = {PLoS ONE},
  month = {nov},
  number = {11},
  pmid = {26571487},
  publisher = {Public Library of Science},
  title = {Hypothetical outcome plots outperform error bars and violin plots for inferences about reliability of variable ordering},
  volume = {10},
  year = {2015}
}

@article{Kale2021,
  author = {Kale, Alex and Kay, Matthew and Hullman, Jessica},
  doi = {10.1109/TVCG.2020.3030335},
  eprint = {2007.14516},
  issn = {19410506},
  journal = {IEEE Transactions on Visualization and Computer Graphics},
  number = {2},
  pages = {272--282},
  pmid = {33048681},
  title = {Visual reasoning strategies for effect size judgments and decisions},
  volume = {27},
  year = {2021}
}

@article{Maceachren2012,
  author = {Maceachren, Alan M. and Roth, Robert E. and O'Brien, James and Li, Bonan and Swingley, Derek and Gahegan, Mark},
  doi = {10.1109/TVCG.2012.279},
  issn = {10772626},
  journal = {IEEE Transactions on Visualization and Computer Graphics},
  number = {12},
  pages = {2496--2505},
  publisher = {IEEE},
  title = {Visual semiotics \& uncertainty visualization: An empirical study},
  volume = {18},
  year = {2012}
}

@article{Padilla2021,
  author = {Padilla, Lace and Powell, Maia and Kay, Matthew and Hullman, Jessica},
  doi = {10.3389/fpsyg.2020.579267},
  issn = {16641078},
  journal = {Frontiers in Psychology},
  month = {jan},
  publisher = {Frontiers Media S.A.},
  title = {Uncertain About Uncertainty: How Qualitative Expressions of Forecaster Confidence Impact Decision-Making With Uncertainty Visualizations},
  volume = {11},
  year = {2021}
}

@article{Vanderplas2015,
  title={Signs of the sine illusion—why we need to care},
  author={VanderPlas, Susan and Hofmann, Heike},
  journal={Journal of Computational and Graphical Statistics},
  volume={24},
  number={4},
  pages={1170--1190},
  year={2015},
  publisher={Taylor \& Francis}
}

@article{vanderplas2017,
  title={Clusters beat trend!? {T}esting feature hierarchy in statistical graphics},
  author={VanderPlas, Susan and Hofmann, Heike},
  journal={Journal of Computational and Graphical Statistics},
  volume={26},
  number={2},
  pages={231--242},
  year={2017},
  publisher={Taylor \& Francis}
}

@article{Hofmann2012,
  title = {Graphical Tests for Power Comparison of Competing Designs},
  author = {Hofmann, Heike and Follett, Lendie and Majumder, Mahbubul and Cook, Dianne},
  year = {2012},
  month = dec,
  journal = {IEEE Transactions on Visualization and Computer Graphics},
  volume = {18},
  number = {12},
  pages = {2441--2448},
  issn = {1077-2626},
  doi = {10.1109/TVCG.2012.230},
  urldate = {2022-12-22},
  langid = {english}
}

@article{Patrick2023,
  title = {A Plot Is Worth a Thousand Tests: Assessing Residual Diagnostics with the Lineup Protocol},
  shorttitle = {A Plot Is Worth a Thousand Tests},
  author = {Li, Weihao and Cook, Dianne and Tanaka, Emi and VanderPlas, Susan},
  year = {2024},
  month = may,
  journal = {Journal of Computational and Graphical Statistics},
  pages = {1--19},
  issn = {1061-8600, 1537-2715},
  doi = {10.1080/10618600.2024.2344612},
  urldate = {2024-09-24},
  langid = {english}
}

@article{Padilla2017,
  author = {Padilla, Lace and Ruginski, Ian and Creem-Regehr, Sarah},
  doi = {10.1186/s41235-017-0076-1},
  issn = {23657464},
  journal = {Cognitive Research: Principles and Implications},
  month = {dec},
  number = {1},
  publisher = {Springer},
  title = {Effects of ensemble and summary displays on interpretations of geospatial uncertainty data},
  volume = {2},
  year = {2017}
}

@article{Correll2014,
  author = {Correll, Michael and Gleicher, Michael},
  doi = {10.1109/TVCG.2014.2346298},
  issn = {10772626},
  journal = {IEEE Transactions on Visualization and Computer Graphics},
  number = {12},
  pages = {2142--2151},
  pmid = {26356928},
  publisher = {IEEE},
  title = {Error bars considered harmful: Exploring alternate encodings for mean and error},
  volume = {20},
  year = {2014}
}

@article{Cheong2016,
  author = {Cheong, Lisa and Bleisch, Susanne and Kealy, Allison and Tolhurst, Kevin and Wilkening, Tom and Duckham, Matt},
  doi = {10.1080/13658816.2015.1131829},
  issn = {13623087},
  journal = {International Journal of Geographical Information Science},
  number = {7},
  pages = {1377--1404},
  publisher = {Taylor & Francis},
  title = {Evaluating the impact of visualization of wildfire hazard upon decision-making under uncertainty},
  url = {http://dx.doi.org/10.1080/13658816.2015.1131829},
  volume = {30},
  year = {2016}
}

@inproceedings{Ndlovu2023,
  title={Taken By Surprise? Evaluating how Bayesian Surprise \& Suppression Influences Peoples’ Takeaways in Map Visualizations},
  author={Ndlovu, Akim and Shrestha, Hilson and Harrison, Lane T},
  booktitle={2023 IEEE Visualization and Visual Analytics (VIS)},
  pages={136--140},
  year={2023},
  organization={IEEE}
}

@article{Kale2018,
  title={Hypothetical outcome plots help untrained observers judge trends in ambiguous data},
  author={Kale, Alex and Nguyen, Francis and Kay, Matthew and Hullman, Jessica},
  journal={IEEE transactions on visualization and computer graphics},
  volume={25},
  number={1},
  pages={892--902},
  year={2018},
  publisher={IEEE}
}

@article{Blenkinsop2000,
  author = {Blenkinsop, Steve and Fisher, Pete and Bastin, Lucy and Wood, Jo},
  doi = {10.3138/3645-4v22-0m23-3t52},
  issn = {03177173},
  journal = {Cartographica},
  number = {1},
  pages = {1--13},
  title = {Evaluating the perception of uncertainty in alternative visualization strategies},
  volume = {37},
  year = {2000}
}

@misc{wu2023rational,
      title={The Rational Agent Benchmark for Data Visualization}, 
      author={Yifan Wu and Ziyang Guo and Michails Mamakos and Jason Hartline and Jessica Hullman},
      year={2023},
      eprint={2304.03432},
      archivePrefix={arXiv},
      primaryClass={cs.HC}
}

@article{Benjamin2018,
  author = {Benjamin, Daniel M. and Budescu, David V.},
  doi = {10.3389/fpsyg.2018.00403},
  issn = {16641078},
  journal = {Frontiers in Psychology},
  number = {MAR},
  pages = {1--17},
  title = {The role of type and source of uncertainty on the processing of climate models projections},
  volume = {9},
  year = {2018}
}

@article{Kim2019,
  archivePrefix = {arXiv},
  arxivId = {1901.02949},
  author = {Kim, Yea Seul and Walls, Logan A. and Krafft, Peter and Hullman, Jessica},
  doi = {10.1145/3290605.3300912},
  eprint = {1901.02949},
  isbn = {9781450359702},
  journal = {Conference on Human Factors in Computing Systems - Proceedings},
  pages = {1--14},
  title = {A Bayesian cognition approach to improve data visualization},
  year = {2019}
}

@incollection{uncertchap2022,
  author  = {Padilla, Lace and Kay, Matthew and Hullman, Jessica},
  title   = {Uncertainty Visualization},
  booktitle   = {Computational Statistics in Data Science},
  editor    = {Piegorsch, Walter W. and Levine, Richard A. and Zhang, Hao Helen and Lee, Thomas C. M.},
  chapter = {22},
  pages   = {405-426},
  year    = {2022},
  publisher = {John Wiley \& Sons},
  address   = {Hoboken, NJ},
  isbn = {9781119561071}
}

@article{Hullman2021,
  author = {Hullman, Jessica and Gelman, Andrew},
  doi = {10.1162/99608f92.3ab8a587},
  journal = {Harvard Data Science Review},
  pages = {1--70},
  title = {Designing for Interactive Exploratory Data Analysis Requires Theories of Graphical Inference},
  year = {2021}
}

@article{Vanderplas2020,
  title={Testing statistical charts: What makes a good graph?},
  author={Vanderplas, Susan and Cook, Dianne and Hofmann, Heike},
  journal={Annual Review of Statistics and Its Application},
  volume={7},
  number={1},
  pages={61--88},
  year={2020},
  publisher={Annual Reviews}
}

@article{Hullman2019,
  author = {Hullman, Jessica and Qiao, Xiaoli and Correll, Michael and Kale, Alex and Kay, Matthew},
  doi = {10.1109/TVCG.2018.2864889},
  issn = {19410506},
  journal = {IEEE Transactions on Visualization and Computer Graphics},
  month = {jan},
  number = {1},
  pages = {903--913},
  publisher = {IEEE Computer Society},
  title = {In Pursuit of Error: A Survey of Uncertainty Visualization Evaluation},
  volume = {25},
  year = {2019}
}

@article{Tierney2023,
  author = {Tierney, Nicholas and Cook, Dianne},
  doi = {10.18637/jss.v105.i07},
  eprint = {1809.02264},
  issn = {15487660},
  journal = {Journal of Statistical Software},
  number = {7},
  pages = {1--31},
  title = {Expanding Tidy Data Principles to Facilitate Missing Data Exploration, Visualization and Assessment of Imputations},
  volume = {105},
  year = {2023}
}

@article{Franconeri2021,
  author = {Franconeri, Steven L.},
  doi = {10.1177/09637214211009512},
  issn = {14678721},
  journal = {Current Directions in Psychological Science},
  number = {5},
  pages = {367--375},
  title = {Three Perceptual Tools for Seeing and Understanding Visualized Data},
  volume = {30},
  year = {2021}
}

@article{MacEachren1992,
  author = {MacEachren, Alan M},
  journal = {Cartographic Perspectives},
  number = {13},
  pages = {10--19},
  title = {Visualizing Uncertain Information},
  year = {1992},
  month = {Jun},
  url = {https://doi.org/10.14714/CP13.1000},
  doi = {10.14714/CP13.1000}
}

@inproceedings{Begg2014,
    author = {Begg, Steve H. and Welsh, Matthew B. and Bratvold, Reidar B.},
    title = {Uncertainty vs. Variability: What’s the Difference and Why is it Important?},
    volume = {SPE Hydrocarbon Economics and Evaluation Symposium},
    booktitle = {SPE Hydrocarbon Economics and Evaluation Symposium},
    numpages = {21},
    year = {2014},
    month = {05},
    doi = {10.2118/169850-MS},
    url = {https://doi.org/10.2118/169850-MS}
}

@article{Kay2023,
  title={{ggdist}: Visualizations of Distributions and Uncertainty in the Grammar of Graphics},
  author={Kay, Matthew},
  journal={IEEE Transactions on Visualization and Computer Graphics},
  volume={30},
  number={1},
  pages={414--424},
  year={2023},
  publisher={IEEE}
}

@article{Kuhnert2018,
  author = {Kuhnert, P. M. and Pagendam, D. E. and Bartley, R. and Gladish, D. W. and Lewis, S. E. and Bainbridge, Z. T.},
  doi = {10.1071/MF17237},
  issn = {13231650},
  journal = {Marine and Freshwater Research},
  number = {8},
  pages = {1187--1200},
  title = {Making management decisions in the face of uncertainty: A case study using the {Burdekin} catchment in the {Great Barrier Reef}},
  volume = {69},
  year = {2018}
}

@article{Lucchesi2021,
  author = {Lucchesi, Lydia and Kuhnert, Petra and Wikle, Christopher},
  doi = {10.21105/joss.02409},
  journal = {Journal of Open Source Software},
  number = {59},
  pages = {2409},
  title = {{Vizumap}: an {R} package for visualising uncertainty in spatial data},
  volume = {6},
  year = {2021}
}

@article{Smart2019,
  author = {Smart, Stephen and Szafir, Danielle Albers},
  doi = {10.1145/3290605.3300899},
  isbn = {9781450359702},
  journal = {Conference on Human Factors in Computing Systems - Proceedings},
  pages = {1--14},
  title = {Measuring the separability of shape, size, and color in scatterplots},
  year = {2019}
}

@inproceedings{Sarma2024,
  author = {Sarma, Abhraneel and Pu, Xiaoying and Cui, Yuan and Correll, Michael and Brown, Eli T and Kay, Matthew},
  title = {Odds and Insights: Decision Quality in Exploratory Data Analysis Under Uncertainty},
  year = {2024},
  isbn = {9798400703300},
  publisher = {Association for Computing Machinery},
  address = {New York, NY, USA},
  url = {https://doi.org/10.1145/3613904.3641995},
  doi = {10.1145/3613904.3641995},
  booktitle = {Proceedings of the CHI Conference on Human Factors in Computing Systems},
  articleno = {1034},
  numpages = {14},
  location = {Honolulu, HI, USA},
  series = {CHI '24}
}

@misc{Kay2019,
  author = {Kay, Matthew},
  title = {How Much Value Should an Uncertainty Palette Suppress if an Uncertainty Palette Should Suppress Value? {S}tatistical and Perceptual Perspectives},
  year = {2019},
  month = {Oct},
  publisher = {OSF Preprints},
  url = {https://doi.org/10.31219/osf.io/6xcnw},
  doi = {10.31219/osf.io/6xcnw},
}

@article{Goldstein2014,
  author = {Goldstein, Daniel G. and Rothschild, David},
  issn = {19302975},
  journal = {Judgment and Decision Making},
  number = {1},
  pages = {1--14},
  title = {Lay understanding of probability distributions},
  volume = {9},
  year = {2014}
}

@inproceedings{moritz2017trust,
  title={Trust, but verify: Optimistic visualizations of approximate queries for exploring big data},
  author={Moritz, Dominik and Fisher, Danyel and Ding, Bolin and Wang, Chi},
  booktitle={Proceedings of the 2017 CHI conference on human factors in computing systems},
  pages={2904--2915},
  year={2017}
}

@Article{tidyverse,
    title = {Welcome to the tidyverse},
    author = {Hadley Wickham and Mara Averick and Jennifer Bryan and Winston Chang and Lucy {D'Agostino McGowan} and Romain François and Garrett Grolemund and Alex Hayes and Lionel Henry and Jim Hester and Max Kuhn and Thomas Lin Pedersen and Evan Miller and Stephan Milton Bache and Kirill Müller and Jeroen Ooms and David Robinson and Dana Paige Seidel and Vitalie Spinu and Kohske Takahashi and Davis Vaughan and Claus Wilke and Kara Woo and Hiroaki Yutani},
    year = {2019},
    journal = {Journal of Open Source Software},
    volume = {4},
    number = {43},
    pages = {1686},
    doi = {10.21105/joss.01686}
}

@Manual{RColorBrewer,
    title = {{RColorBrewer}: {ColorBrewer} Palettes},
    author = {Erich Neuwirth},
    year = {2022},
    note = {R package version 1.1-3},
    url = {https://CRAN.R-project.org/package=RColorBrewer}
  }

@Manual{scales,
    title = {{scales}: Scale Functions for Visualization},
    author = {Hadley Wickham and Thomas Lin Pedersen and Dana Seidel},
    year = {2023},
    note = {R package version 1.3.0},
    url = {https://CRAN.R-project.org/package=scales}
  }

@Book{sf,
    author = {Edzer Pebesma and Roger Bivand},
    title = {Spatial Data Science: With applications in {R}},
    year = {2023},
    publisher = {Chapman and Hall/CRC},
    url = {https://r-spatial.org/book/},
    doi = {10.1201/9780429459016}
  }

@Manual{urbnmapr,
    title = {urbnmapr: State and county shapefiles in sf and tibble format},
    author = {Sarah Strochak and Kyle Ueyama and Aaron Williams},
    year = {2024},
    note = {R package version 0.0.0.9002},
    url = {https://github.com/UrbanInstitute/urbnmapr}
    }

@Manual{flextable,
    title = {flextable: Functions for Tabular Reporting},
    author = {David Gohel and Panagiotis Skintzos},
    year = {2024},
    note = {R package version 0.9.6},
    url = {https://CRAN.R-project.org/package=flextable}
}

@Article{colorspace,
    title = {Somewhere over the Rainbow: How to Make Effective Use of Colors in Meteorological Visualizations},
    author = {Reto Stauffer and Georg J. Mayr and Markus Dabernig and Achim Zeileis},
    journal = {Bulletin of the American Meteorological Society},
    year = {2009},
    volume = {96},
    number = {2},
    pages = {203--216},
    doi = {10.1175/BAMS-D-13-00155.1}
}

@Manual{rgeos,
    title = {rgeos: Interface to Geometry Engine - Open Source ('GEOS')},
    author = {Roger Bivand and Colin Rundel},
    year = {2023},
    note = {R package version 0.6-3},
    url = {https://CRAN.R-project.org/package=rgeos}
}

@inproceedings{datasaurus, 
  author = {Matejka, Justin and Fitzmaurice, George}, 
  title = {Same Stats, Different Graphs: Generating Datasets with Varied Appearance and Identical Statistics through Simulated Annealing}, 
  year = {2017}, isbn = {9781450346559}, 
  publisher = {Association for Computing Machinery}, 
  address = {New York, NY, USA}, 
  url = {https://doi.org/10.1145/3025453.3025912}, 
  doi = {10.1145/3025453.3025912}, 
  booktitle = {Proceedings of the 2017 CHI Conference on Human Factors in Computing Systems}, 
  pages = {1290–1294}, 
  numpages = {5}, 
  location = {Denver, Colorado, USA}, 
  series = {CHI '17} }

@article{Hadjimichael2024,
	title = {Data visualisation for decision making under deep uncertainty: current challenges and opportunities},
	volume = {19},
	issn = {1748-9326},
	shorttitle = {Data visualisation for decision making under deep uncertainty},
	url = {https://iopscience.iop.org/article/10.1088/1748-9326/ad858b},
	doi = {10.1088/1748-9326/ad858b},
	language = {en},
	number = {11},
	urldate = {2025-11-11},
	journal = {Environmental Research Letters},
	author = {Antonia Hadjimichael and Julius Schlumberger and Marjolijn Haasnoot},
	month = nov,
	year = {2024},
	pages = {111011},
}

@inproceedings{geointerviews,
  title={Uncertainty in Science is Malleable. Advocating for User-Agency in Defining Uncertainty in Visualizations: a Case Study in Geology},
  author={Vanessa Pe{\~n}a-Araya and Consuelo Mart{\'\i}nez Fontaine and Xiang Wei and Guillaume Delpech and Anastasia Bezerianos},
  booktitle={Proceedings of the 2025 CHI Conference on Human Factors in Computing Systems},
  pages={1--18},
  year={2025}
}

@article{ggplot2,
	title = {A Layered Grammar of Graphics},
	volume = {19},
	issn = {10618600},
	url = {https://doi.org/10.1198/jcgs.2009.07098},
	doi = {10.1198/jcgs.2009.07098},
	number = {1},
	journal = {Journal of Computational and Graphical Statistics},
	author = {Wickham, Hadley},
	year = {2010},
	pages = {3--28},
}

@article{MacEachren2005,
	title = {Visualizing geospatial information uncertainty: {What} we know and what we need to know},
	volume = {32},
	issn = {15230406},
	doi = {10.1559/1523040054738936},
	number = {3},
	journal = {Cartography and Geographic Information Science},
	author = {Alan M. MacEachren and Anthony Robinson and Susan Hopper and Steven Gardner and Robert Murray and Mark Gahegan and Elisabeth Hetzler},
	year = {2005},
	note = {ISBN: 1523040054738},
	pages = {139--160},
}

@Manual{distributional,
    title = {{distributional}: Vectorised Probability Distributions},
    author = {Mitchell O'Hara-Wild and Matthew Kay and Alex Hayes and Rob Hyndman},
    year = {2024},
    note = {R package version 0.5.0},
    url = {https://CRAN.R-project.org/package=distributional},
    doi = {10.32614/CRAN.package.distributional}
  }

@article{Brennen2018,
	title = {An instrument for evaluating uncertainty visualization techniques},
	volume = {2018-April},
	doi = {10.1145/3170427.3188649},
	journal = {Conference on Human Factors in Computing Systems - Proceedings},
	author = {Brennen, Andrea and Tuerk, Stephanie},
	year = {2018},
	note = {ISBN: 9781450356206},
	pages = {1--6},
}

@article{Reda2021,
	title = {Rainbows Revisited: Modeling Effective Colormap Design for Graphical Inference},
	volume = {27},
	issn = {1077-2626, 1941-0506, 2160-9306},
	shorttitle = {Rainbows Revisited},
	url = {https://ieeexplore.ieee.org/document/9222327/},
	doi = {10.1109/TVCG.2020.3030439},
	language = {en},
	number = {2},
	urldate = {2026-02-23},
	journal = {IEEE Transactions on Visualization and Computer Graphics},
	author = {Reda, Khairi and Szafir, Danielle Albers},
	month = feb,
	year = {2021},
	pages = {1032--1042},
}

@inproceedings{Lee2021,
  title={Viral visualizations: How coronavirus skeptics use orthodox data practices to promote unorthodox science online},
  author={Lee, Crystal and Yang, Tanya and Inchoco, Gabrielle D and Jones, Graham M and Satyanarayan, Arvind},
  booktitle={Proceedings of the 2021 CHI conference on human factors in computing systems},
  pages={1--18},
  year={2021}
}

@article{Savelli2013,
  title={The advantages of predictive interval forecasts for non-expert users and the impact of visualizations},
  author={Savelli, Sonia and Joslyn, Susan},
  journal={Applied Cognitive Psychology},
  volume={27},
  number={4},
  pages={527--541},
  year={2013},
  publisher={Wiley Online Library}
}

@book{Tukey1977,
  title={Exploratory data analysis},
  author={Tukey, John Wilder and others},
  volume={2},
  year={1977},
  publisher={Springer}
}

@article{Guo2024,
  title={{VMC}: A Grammar for Visualizing Statistical Model Checks},
  author={Ziyang Guo and Alex Kale and Matthew Kay and Jessica Hullman},
  journal={IEEE Transactions on Visualization and Computer Graphics},
  year={2025},
  publisher = {IEEE Educational Activities Department}, 
  url = {https://doi.org/10.1109/TVCG.2024.3456402}, 
  doi = {10.1109/TVCG.2024.3456402},
  volume = {31}, 
  number = {1},
  pages = {798–808}, 
  numpages = {11}
}

@inproceedings{Savvides2019,
	address = {Anchorage AK USA},
	title = {Significance of Patterns in Data Visualisations},
	isbn = {978-1-4503-6201-6},
	url = {https://doi.org/10.1145/3292500.3330994}, 
	doi = {10.1145/3292500.3330994},
	language = {en},
	booktitle = {Proceedings of the 25th ACM SIGKDD International Conference on Knowledge Discovery \& Data Mining},
	publisher = {Association for Computing Machinery}, 
	author = {Rafael Savvides and Andreas Henelius and Emilia Oikarinen and Kai Puolamäki},
	month = jul,
	year = {2019},
	pages = {1509--1517},
}

@article{Majumder2013,
	title = {Validation of Visual Statistical Inference, Applied to Linear Models},
	volume = {108},
	issn = {0162-1459, 1537-274X},
	url = {http://www.tandfonline.com/doi/abs/10.1080/01621459.2013.808157},
	doi = {10.1080/01621459.2013.808157},
	language = {en},
	number = {503},
	urldate = {2025-11-11},
	journal = {Journal of the American Statistical Association},
	author = {Majumder, Mahbubul and Hofmann, Heike and Cook, Dianne},
	month = sep,
	year = {2013},
	pages = {942--956},
}

@article{Chakraborty2024,
	title = {The influence of uncertainty visualization on cognitive load in a safety- and time-critical decision-making task},
	volume = {38},
	issn = {1365-8816, 1362-3087},
	url = {https://www.tandfonline.com/doi/full/10.1080/13658816.2024.2348747},
	doi = {10.1080/13658816.2024.2348747},
	language = {en},
	number = {8},
	urldate = {2025-11-11},
	journal = {International Journal of Geographical Information Science},
	author = {Chakraborty, Suvodip and Kiefer, Peter and Raubal, Martin},
	month = aug,
	year = {2024},
	pages = {1583--1610}
	}

@article{Sarma2023,
	title = {Evaluating the Use of Uncertainty Visualisations for Imputations of Data Missing at Random in Scatterplots},
	volume = {29},
	issn = {19410506},
	doi = {10.1109/TVCG.2022.3209348},
	number = {1},
	journal = {IEEE Transactions on Visualization and Computer Graphics},
	author = {Sarma, Abhraneel and Guo, Shunan and Hoffswell, Jane and Rossi, Ryan and Du, Fan and Koh, Eunyee and Kay, Matthew},
	year = {2023},
	pages = {602--612},
}

@article{Koonchanok2023,
	title = {Visual Belief Elicitation Reduces the Incidence of False Discovery},
	doi = {10.1145/3544548.3580808},
	journal = {Conference on Human Factors in Computing Systems - Proceedings},
	author = {Koonchanok, Ratanond and Tawde, Gauri Yatindra and Narayanasamy, Gokul Ragunandhan and Walimbe, Shalmali and Reda, Khairi},
	year = {2023},
}

@article{Padilla2022,
	title = {Impact of {COVID}-19 forecast visualizations on pandemic risk perceptions},
	volume = {12},
	issn = {2045-2322},
	url = {https://www.nature.com/articles/s41598-022-05353-1},
	doi = {10.1038/s41598-022-05353-1},
	number = {1},
	urldate = {2022-04-15},
	journal = {Scientific Reports 2022 12:1},
	publisher = {Nature Publishing Group},
	author = {Padilla, Lace and Hosseinpour, Helia and Fygenson, Racquel and Howell, Jennifer and Chunara, Rumi and Bertini, Enrico},
	month = feb,
	year = {2022},
	pages = {1--14}
	}

@article{Yang2023,
  title={Swaying the public? {I}mpacts of election forecast visualizations on emotion, trust, and intention in the 2022 us midterms},
  author={Yang, Fumeng and Cai, Mandi and Mortenson, Chloe and Fakhari, Hoda and Lokmanoglu, Ayse D and Hullman, Jessica and Franconeri, Steven and Diakopoulos, Nicholas and Nisbet, Erik C and Kay, Matthew},
  journal={IEEE Transactions on Visualization and Computer Graphics},
  volume={30},
  number={1},
  pages={23--33},
  year={2023},
  publisher={IEEE}
}

@Manual{ggdibbler,
    title = {ggdibbler: Add Uncertainty to Data Visualisations},
    author = {Harriet Mason and Dianne Cook and Sarah Goodwin and Susan VanderPlas},
    year = {2026},
    url = {https://github.com/harriet-mason/ggdibbler},
  }

@Manual{patchwork,
    title = {{patchwork}: The Composer of Plots},
    author = {Thomas Lin Pedersen},
    year = {2025},
    note = {R package version 1.3.2},
    url = {https://CRAN.R-project.org/package=patchwork},
    doi = {10.32614/CRAN.package.patchwork},
  }

@Manual{ggthemes,
    title = {ggthemes: Extra Themes, Scales and Geoms for 'ggplot2'},
    author = {Jeffrey B. Arnold},
    year = {2024},
    note = {R package version 5.1.0},
    url = {https://CRAN.R-project.org/package=ggthemes},
    doi = {10.32614/CRAN.package.ggthemes},
}

@article{mtcars,
  author = {Henderson and Velleman},
  year = 1981, 
  title = {Building multiple regression models interactively},
  journal = {Biometrics}, 
  volume = 37, 
  pages = {391-411}
}

@article{huebner2016,
  title = {A systematic approach to initial data analysis is good research practice},
  journal = {The Journal of Thoracic and Cardiovascular Surgery},
  volume = {151},
  number = {1},
  pages = {25-27},
  year = {2016},
  issn = {0022-5223},
  doi = {10.1016/j.jtcvs.2015.09.085},
  url = {https://doi.org/10.1016/j.jtcvs.2015.09.085},
  author = {Marianne Huebner and Werner Vach and Saskia {le Cessie}},
}

@article{chatfield,
 ISSN = {00359238, 23972327},
 doi = {10.2307/2981969},
 URL = {https://doi.org/10.2307/2981969},
 author = {C. Chatfield},
 journal = {Journal of the Royal Statistical Society. Series A (General)},
 number = {3},
 pages = {214--253},
 publisher = {[Royal Statistical Society, Oxford University Press]},
 title = {The Initial Examination of Data},
 urldate = {2026-03-19},
 volume = {148},
 year = {1985}
}

@book{bkw1980,
  title={Regression Diagnostics: Identifying Influential Data and Sources of Collinearity},
  author={Belsley, D.A. and Kuh, E. and Welsch, R.E.},
  isbn={9780471058564},
  lccn={lc79019876},
  series={Wiley Series in Probability and Statistics},
  url={https://onlinelibrary.wiley.com/doi/book/10.1002/0471725153},
  doi = {10.1002/0471725153},
  year={1980},
  address = {New York},
  publisher={Wiley}
}

@article{simons1997,
  title = {Change blindness}, 
  author = {D. J. Simons and D. T. Levin},
  year = 1997, 
  journal = {Trends in Cognitive Sciences},
  volume = {1},
  issue = {7},
  pages = {261–267},
  doi = {10.1016/S1364-6613(97)01080-2},
  url = {https://doi.org/10.1016/S1364-6613(97)01080-2}
}

@book{fpp3,
  title = {Forecasting: principles and practice, 3rd edition}, 
  author = {Hyndman, R.J. and Athanasopoulos, G},
  year = {2021},
  publisher = {OTexts},
  address = {Melbourne, Australia},
  url = {https://otexts.com/fpp3/}
}

@inproceedings{heer2010, 
  author = {Heer, Jeffrey and Bostock, Michael}, 
  title = {Crowdsourcing graphical perception: using mechanical turk to assess visualization design}, 
  year = {2010}, 
  isbn = {9781605589299}, 
  publisher = {Association for Computing Machinery}, 
  address = {New York, NY, USA}, 
  url = {https://doi.org/10.1145/1753326.1753357}, 
  doi = {10.1145/1753326.1753357}, 
  booktitle = {Proceedings of the SIGCHI Conference on Human Factors in Computing Systems}, 
  pages = {203–212}, 
  numpages = {10}, 
  location = {Atlanta, Georgia, USA}, 
  series = {CHI '10} 
}

@Manual{broom,
    title = {{broom}: Convert Statistical Objects into Tidy Tibbles},
    author = {David Robinson and Alex Hayes and Simon Couch and Emil Hvitfeldt},
    year = {2026},
    note = {R package version 1.0.12},
    url = {https://CRAN.R-project.org/package=broom},
    doi = {10.32614/CRAN.package.broom},
  }

\end{document}